\documentclass[letterpaper,twocolumn,10pt]{article}
\makeatletter\let\oldMakeTitle\@maketitle\makeatother

\usepackage{usenix2019_v3}

\usepackage{tikz}
\usepackage{changepage}
\usepackage{amsmath}
\usepackage[colorinlistoftodos]{todonotes}
\usepackage{titlesec}
\usepackage{cite}
\usepackage{xspace}
\usepackage{tabularx}
\usepackage{verbatim} 
\usepackage{multirow}
\usepackage{subfig}
\usepackage{csquotes}
\usepackage{placeins}
\usepackage{array}
\usepackage{numprint}
\usepackage{boxhandler}
\usepackage{afterpage}
\usepackage{wrapfig}
\usepackage{amsfonts}
\usepackage{datetime}
\usepackage{amsthm}
\usepackage{graphics}

\usepackage{algorithm}
\usepackage{nicefrac}
\usepackage{xcolor}
\usepackage{authblk}
\usepackage{varwidth}
\usepackage{sidecap}
\usepackage{graphicx,calc}
\usepackage[noend]{algpseudocode} 
\algrenewcommand\alglinenumber[1]{\tiny #1:}
\usepackage{soul}
\usepackage{fancyhdr}

\newif\ifsubmission
\submissiontrue 
\newif\ifrevision
\revisionfalse 
\newif\ifminorrevision
\minorrevisionfalse

\newcommand{\ourname}{\mbox{\textsc{FLAME}}\xspace}
\newcommand{\ournameGen}{\mbox{\textsc{FLAME}'s}\xspace}

\newcommand{\diot}{D\"IoT \xspace}
\newcommand{\thrdfed}{3DFed \xspace}

\newcommand{\etal}{\emph{et~al.}}

\newcommand{\sota}{state-of-the-art }
\newcommand{\ie}{i.e.}

\newcommand{\iid}{iid\xspace}
\newcommand{\noniid}{non-iid\xspace}

\newcommand{\adversaryClient}{{\ensuremath{\mathcal{A}}}}
\newcommand{\adversaryServer}{{\ensuremath{\mathcal{A}^s}}}

\newcommand{\verfiveChanged}[1]{\textcolor{black}{#1}}

\newtheorem{theorem}{Theorem}

\newcommand{\DataFLIoTBenign}{\textit{\ourname-Benign}\xspace}
\newcommand{\DataDIoTBenign}{\textit{DIoT-Benign}\xspace}
\newcommand{\DataUNSWBenign}{\textit{UNSW-Benign}\xspace}
\newcommand{\DataDIoTAttack}{\textit{DIoT-Attack}\xspace}

\newcommand{\reddit}{Reddit\xspace}
\newcommand{\cifar}{CIFAR-10\xspace}

\newcommand{\mnist}{MNIST\xspace}
\newcommand{\tinyImage}{Tiny-ImageNet\xspace}

\newcommand{\lnorm}{\text{L$_2$-norm}\xspace}
\newcommand{\lnorms}{\text{L$_2$-norms}\xspace}

\newcommand{\dist}{dist}

\usepackage{xcolor}

\newdateformat{mydateformat}{\THEYEAR-\THEMONTH-\THEDAY}
\newcommand{\hdbscan}{\mbox{HDBSCAN}\xspace}
\newcommand{\dbscan}{\mbox{DBSCAN}\xspace}
\newcommand{\kmeans}{\mbox{K-means}\xspace}

\newtheorem{definition}{Definition} 

\newcommand{\mcrot}[4]{\multicolumn{#1}{#2}{\rlap{\rotatebox{#3}{#4}~}}} 
\newcommand{\bul}{\textbullet}
\newcommand{\cir}{$\circ$}

\newcommand{\ModelFiltering}{\textit{Model Filtering}\xspace}

\newcommand{\Modelreplacement}{\emph{Model-replacement}\xspace}
\newcommand{\modelreplacement}{\emph{model-replacement}\xspace}

\newcommand{\Constrainandscale}{\emph{Constrain-and-scale}\xspace}
\newcommand{\constrainandscale}{\emph{constrain-and-scale}\xspace}

\newcommand{\edgeAttack}{Edge-Case\xspace}
\newcommand{\edgePGDAttack}{Edge-Case PGD\xspace}
\newcommand{\degNonIID}{\ensuremath{\text{Deg}_{\text{nIID}}}}

\newcommand{\trainandscale}{\emph{train-and-scale}\xspace}

\newcommand{\gAccuracy}{\textit{Impact}\xspace}
\newcommand{\gStealthiness}{\textit{Stealthiness}\xspace}

\newcommand{\numberofDeviceType}{24\xspace}
\newcommand{\numberofDevice}{78\xspace}
\newcommand{\numberofAttackType}{13\xspace}

\newcommand{\sect}{§}

\newcommand{\equ}{Eq.}

\newcommand{\iotTraffic}{IoT-Traffic}
\newlength\myheight
\newlength\mydepth
\settototalheight\myheight{Xygp}
\definecolor{darkgreen}{rgb}{0.0, 0.2, 0.13}
\definecolor{darkspringgreen}{rgb}{0.09, 0.45, 0.27}
\definecolor{forestgreen}{rgb}{0.13, 0.55, 0.13}

\ifminorrevision
    \newcommand{\minorRevision}[2]{{\color{blue}{#1}}}
    
    \newcommand{\addressedComment}[1]{\todo{\footnotesize{Addresses #1}}} 
\else
    \newcommand{\minorRevision}[2]{{\color{black}{#1}}}
    \newcommand{\addressedComment}[1]{}
\fi

\ifrevision
    \newcommand{\revisionChanged}[2]{{\color{blue}{#1}}}
    \newcommand{\revisionAdded}[2]{{\color{blue}{#1}}}

\else
    \newcommand{\revisionChanged}[2]{#1}
    \newcommand{\revisionAdded}[2]{#1}
\fi

\ifsubmission
	\newcommand{\TODO}[1]{}
	
	\newcommand{\HCHANGED}[1]{#1}
	
	\newcommand{\TOASK}[1]{}
	\newcommand{\tsc}[1]{}
	\newcommand{\ALL}[1]{}

\else 
	\newcommand{\TODO}[1]{\marginpar{\textcolor{red}{TODO: #1}}}
	
	\newcommand{\HCHANGED}[1]{\textcolor{black}{#1}}

	\newcommand{\TOASK}[1]{\marginpar{\textcolor{green}{?: #1}}}
	\newcommand{\ALL}[1]{\textcolor{green}{ALL: #1}}

\fi

\newcommand{\ma}{\mbox{\ensuremath{\mathit{MA}}}\xspace}
\newcommand{\ba}{\mbox{\ensuremath{\mathit{BA}}}\xspace}
\newcommand{\pmr}{\mbox{\ensuremath{\mathit{PMR}}}\xspace}
\newcommand{\pmrs}{\mbox{\ensuremath{\mathit{PMRs}}}\xspace}
\newcommand{\pdr}{\mbox{\ensuremath{\mathit{PDR}}}\xspace}
\newcommand{\tpr}{\mbox{\ensuremath{\mathit{TPR}}}\xspace}
\newcommand{\tnr}{\mbox{\ensuremath{\mathit{TNR}}}\xspace}

\newcommand{\EqualTableFontSize}[1]{\footnotesize{#1}}

\newcommand{\mypar}[1]{\noindent{}\textbf{#1.}}

\makeatletter
\renewcommand\AB@affilsepx{; \protect\Affilfont}
\makeatother

\usepackage{filecontents}

\begin{document}

\date{}

\title{\vspace{-5em}  \Large \bf \ourname: Taming Backdoors in Federated Learning (Extended Version 1)}

\author[1]{Thien Duc Nguyen}
\author[1]{Phillip Rieger}
\author[2]{Huili Chen}
\author[1]{Hossein Yalame} 
\author[1]{Helen Möllering} 
\author[1]{\\Hossein Fereidooni}
\author[3]{Samuel Marchal}
\author[1]{Markus Miettinen}
\author[4]{Azalia Mirhoseini}
\author[1]{\\ Shaza Zeitouni} 
\author[2]{Farinaz Koushanfar} 
\author[1]{Ahmad-Reza Sadeghi}
\author[1]{Thomas Schneider \vspace{-0.3cm}}

\affil[1]{\textit{Technical University of Darmstadt, Germany}\thanks{Emails: \{ducthien.nguyen, ahmad.sadeghi\}@trust.tu-darmstadt.de}} 


\affil[2]{\textit{University of California San Diego, USA}}
\affil[3]{\textit{Aalto University and F-Secure, Finland}}
\affil[4]{\textit{Google, USA \vspace{-2 cm}}}

\maketitle


\begin{abstract}
\revisionChanged{Federated Learning (FL) is a collaborative machine learning approach allowing participants to jointly train a model without sharing their private, potentially sensitive local datasets with others. Despite its benefits, FL is vulnerable to so-called \emph{backdoor attacks},}{} in which an adversary injects manipulated model updates into the federated model aggregation process so that the resulting model will provide targeted false predictions for specific adversary-chosen inputs. \revisionChanged{Proposed defenses against backdoor attacks}{} based on detecting and filtering out malicious model updates consider only very specific and limited attacker models, whereas defenses based on differential privacy-inspired noise injection significantly deteriorate the benign performance of the aggregated model.
To address these deficiencies, we introduce \ourname, a defense framework that \revisionChanged{estimates the sufficient amount of noise to be injected to ensure the elimination of backdoors. To minimize the required amount of noise, \ourname uses a model clustering and weight clipping approach. This ensures that \ourname can maintain the benign performance of the aggregated model while effectively eliminating adversarial backdoors.}{} Our evaluation of \ourname on several datasets stemming from application areas, including image classification, word prediction, and IoT intrusion detection, demonstrates that \ourname removes backdoors effectively with a negligible impact on the benign performance of the models. 

\verfiveChanged{ Furthermore, following the considerable attention that our research has received after its presentation at USENIX SEC 2022, FLAME has become the subject of numerous investigations proposing diverse attack methodologies in an attempt to circumvent it. As a response to these endeavors, we provide a comprehensive analysis of these attempts. Our findings show that these papers (e.g., 3DFed \cite{li2023sp}) have not fully comprehended nor correctly employed the fundamental principles underlying FLAME. Moreover, their evaluations appear to be incomplete, with handpicked results chosen selectively. Consequently, in succinct terms, our defense mechanism, FLAME, effectively repels these attempted attacks.}

\end{abstract}


\vspace{-0.5em}
\section{Introduction}
\label{sec:intro}
\vspace{-0.8em}
\emph{Federated learning} (FL) is an emerging collaborative machine learning trend with many applications, such as next-word prediction for mobile keyboards~\cite{mcmahan2017googleGboard}, medical imaging~\cite{sheller}, and intrusion detection for IoT~\cite{nguyen2019diot} to name a few. In FL, clients locally train models based on local training data and then provide these model updates to a central aggregator that combines them into a global model. The global model is then propagated back to the clients \mbox{for the next training iteration.}

FL promises efficiency and scalability as the training is distributed among many clients and executed in parallel. 
In particular, FL improves privacy by enabling clients to keep their training data locally~\cite{mcmahan2017aistatsCommunication}.

Despite its benefits, FL has been shown to be vulnerable to so-called \emph{poisoning attacks} where the adversary manipulates the local models of a subset of clients participating in the federation so that the malicious updates get aggregated into the global model. Untargeted poisoning attacks merely aim at deteriorating the performance of the global model and can be defeated by validating the performance of uploaded models~\cite{cao2020fltrust}. In this paper, we, therefore, focus on the more challenging problem of \emph{backdoor attacks}~\cite{bagdasaryan,nguyen2020diss,xie2020dba, wang2020attack}, 
i.e., targeted poisoning attacks in which the adversary seeks to stealthily manipulate the resulting global model in a way that attacker-controlled inputs result in incorrect predictions chosen by the adversary.\\

\vspace{-1.3em}
\noindent\textbf{Deficiencies of existing defenses.}
Existing defenses against backdoor attacks can be roughly divided into two categories: The first one comprises anomaly detection-based approaches~\cite{shen,blanchard, fung, andreina2020baffle} for identifying and removing potentially poisoned model updates. However, these solutions are effective only under very specific adversary models, as they make detailed assumptions about the attack strategy of the adversary and/or the underlying distribution of the benign or adversarial datasets. If these very specific assumptions do not hold, the defenses may fail.
The second category is inspired by differential privacy (DP) techniques~\cite{sun2019can, bagdasaryan}, where individual weights\footnote{Parameters of neural network models typically consist of 'weights' and 'biases'. For the purposes of this paper, however, these parameters can be treated identically and we will refer to them as 'weights' for brevity.} of model updates are clipped to a maximum threshold and random noise is added to the weights for diluting/reducing the impact of potentially poisoned model updates on the aggregated global model. In contrast to the first category, DP techniques~\cite{sun2019can, bagdasaryan} are applicable in a generic adversary model without specific assumptions about adversarial behavior and data distributions and are effective in eliminating the impact of malicious model updates. However, straightforward application of DP approaches severely deteriorates the benign performance of the aggregated model because the amount of noise required to ensure effective elimination of backdoors also results in significant modifications of individual weights of benign model updates~\cite{bagdasaryan, wang2020attack}.

In this paper, we develop a resilient defense against backdoors by combining the benefits of both defense types without suffering from the limitations (narrow attacker model, assumptions about data distributions) and drawbacks (loss of benign performance) of existing approaches. To this end, we introduce an approach in which the detection of anomalous model updates and tuned clipping of weights are combined to minimize the amount of noise needed for backdoor removal of the aggregated model while preserving its benign performance. 

\vspace{0.2em}
\noindent\textbf{Our Goals and Contributions.} 
\label{contribution}
We present \emph{\ourname}, a resilient aggregation framework for FL that eliminates the impact of backdoor attacks while maintaining the benign performance of the aggregated model. This is achieved by three modules: DP-based noising of model updates to remove backdoor contributions, automated model clustering approach to identify and eliminate potentially poisoned model updates, and model weight clipping before aggregation to limit the impact of malicious model updates on the aggregation result. 
The last two modules can significantly reduce the amount of random noise required by DP noising for backdoor elimination. In particular, our contributions are as follows:
\begin{itemize}
    \vspace{-0.6em}
   \item We present \ourname, a defense framework against backdoor attacks in FL that is capable of eliminating backdoors without impacting the benign performance of the aggregated model. Contrary to earlier backdoor defenses, \ourname is applicable in a \emph{generic} adversary model, i.e., it does not rely on strong assumptions about the attack strategy of the adversary, nor about the underlying data distributions of benign and adversarial datasets (\sect\ref{sec:high-level}).
   \vspace{-0.3em}
    \item We show that the amount of required Gaussian noise can be radically reduced by: a) applying our clustering approach to remove potentially malicious model updates and b) clipping the weights of local models at a proper level to constrain the impact of individual (especially malicious) models on the aggregated model.
       (\sect\ref{sec:defense})
    
    \vspace{-0.3em}   
    \item We provide a noise boundary proof for the amount of Gaussian noise required by noise injection (inspired by DP) to eliminate backdoor contributions (\sect\ref{sec:security-analysis}).
    
    \vspace{-0.3em}
    \item We extensively evaluate our defense framework on real-world datasets from three very different application areas. We show that \ourname reduces the amount of required noise so that the benign performance of the aggregated model does not degrade significantly, providing a crucial advantage over state-of-the-art defenses using a straightforward injection of DP-based noise (\sect\ref{sec:exp-result}).
    \vspace{-0.3em}
    \item \verfiveChanged{ We revisit recent attacks claiming to undermine FLAME \cite{Xu2022ACSAC, li2023sp}. Our analysis establishes that these attempts are flawed, failing to take into account the nuances of FLAME's defense mechanism and, regrettably, propagating inaccuracies, as elaborated upon in Section \ref{sec:new-attacks}.}

\end{itemize}
\vspace{-0.5em}
As an orthogonal aspect, we also consider how the privacy of model updates against an honest-but-curious aggregator can be preserved and develop a secure multi-party computation approach that can preserve the privacy of individual model updates while realizing our backdoor defense approach (\sect\ref{sec:privacy}).

\vspace{-0.5em}
\section{Background and Problem Setting}
\label{sec:background}
\vspace{-1em}
\subsection{Federated Learning}
\label{sec:bg_fl}
\vspace{-0.5em}
Federated Learning~\cite{mcmahan2017aistatsCommunication, sheller2018medical} is a concept for distributed machine learning that links $n$ clients and an aggregator to collaboratively build a global model $G$.
\revisionChanged{ In a training iteration $ t \in \{1,\ldots, T\} $, each client $i \in \{1, \ldots, n\} $ locally trains a local model $W_i$ with $p$ parameters (indicating both weights and biases) $w_i^1, \ldots, w_i^p$ based on the previous global model $G_{t-1}$ using its local data $D_i$ and sends it to the aggregator which aggregates the received models $W_i$ into the global model $G_t$.}{}

Several aggregation mechanisms have been proposed recently: 1) \textit{Federated Averaging} (FedAvg)~\cite{mcmahan2017aistatsCommunication}, 2) \emph{Krum}~\cite{blanchard}, 3) \emph{Adaptive Federated Averaging}~\cite{munoz}, and 4) \emph{Trimmed mean or median}~\cite{yin2018byzantine}.
 Although we evaluate \ourname's effectiveness on several aggregation mechanisms in \sect\ref{sec:eval-flguard}, we generally focus on FedAvg in this work as it is commonly applied in FL \cite{mcmahan2017googleGboard,sheller2018medical, nguyen2019diot, ren2019edgeIoT,smith2017nipsMultitask,huang2019arxivMedical,fereidooni2021safelearn,fereidooni2022fedcri} and related work on backdoor attacks~\cite{shen,fung, bagdasaryan,xie2020dba, wang2020attack}. In FedAvg, the global model is updated by averaging the weighted models as follows:
$G_{t} = \Sigma_{i=1}^n \nicefrac{s_i\times W_i}{s}, \text{ where }  s_i = \Vert D_i \Vert, s = \Sigma_{i=1}^n s_i.$
However, in practice, a malicious client might provide falsified information about its dataset size (i.e., a large number) to amplify the relative weight of its updates \cite{wang2020attack}. Previous works often employed equal weights ($s_i= 1/n$) for the contributions of all clients~\cite{bagdasaryan,xie2020dba, shen}. We adopt this approach in this paper, i.e., we set $G_{t} = \Sigma_{i=1}^n \nicefrac{W_i}{n}$. \revisionAdded{Further, other state-of-the-art aggregation rules, e.g., \emph{Krum}~\cite{blanchard}, \emph{Adaptive Federated Averaging}~\cite{munoz}, and \emph{Trimmed mean or median}~\cite{yin2018byzantine} also do not consider the sizes of local training datasets by design.}

\vspace{-1.2em}
\subsection{Backdoor Attacks on Federated Learning}
\label{sec:existing-attacks}
\vspace{-0.4em}

\begin{figure}[t]
	\centering
	\setlength{\belowcaptionskip}{-8pt}
	\includegraphics[width=0.65\columnwidth]{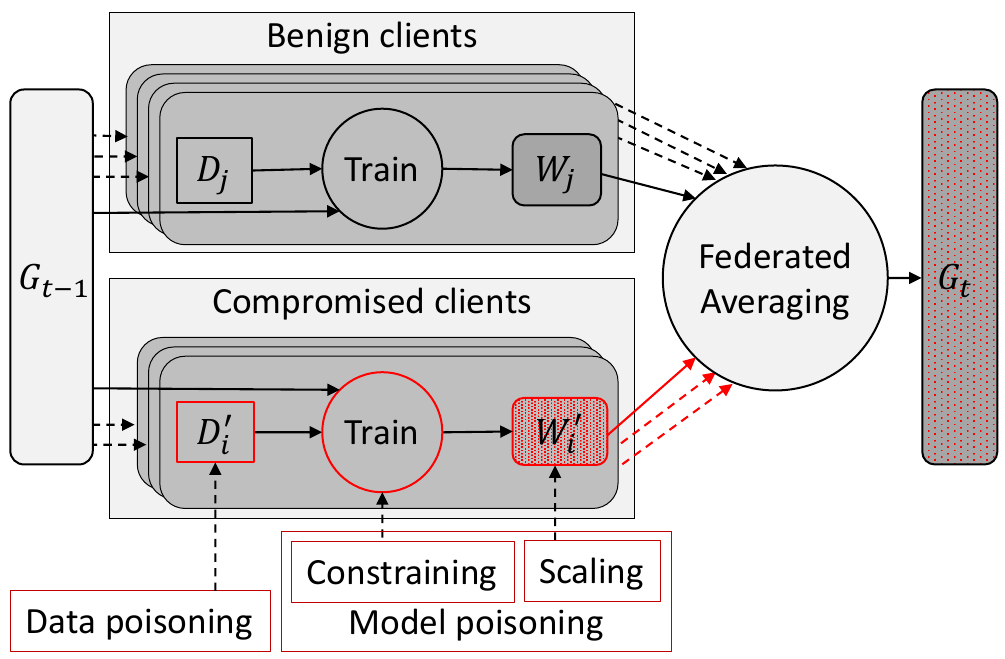}
	\vspace{-0.7em}
	\caption{An overview of backdoor attacks. }
	\label{fig:backdoor-attack}
	\vspace{-0.7em}
\end{figure}

In backdoor attacks, the adversary \adversaryClient\xspace manipulates the local models $W_i$ of $k$ compromised clients to obtain poisoned models $W'_i$ that are then aggregated into the global model $ G_t $ and thus affect its properties. In particular, ~$\adversaryClient$ wants the poisoned model $ G_t' $ to behave normally on all inputs except for specific attacker-chosen inputs $x \in I_\adversaryClient$ (where $I_\adversaryClient$ denotes the so-called \emph{trigger set}) for which attacker-chosen (incorrect) predictions should be output.
Figure \ref{fig:backdoor-attack} shows common techniques used in FL backdoor attacks, including 1) data poisoning, e.g.,~\cite{shen, nguyen2020diss, xie2020dba}, where $\adversaryClient$ manipulates training datasets of models, and 2) model poisoning, e.g.,~\cite{bagdasaryan, wang2020attack} where $\adversaryClient$ manipulates the training process or the trained models themselves. 
Next, we will briefly discuss these attack techniques.

\vspace{0.2em}
\noindent\textbf{Data Poisoning.} 
\revisionChanged{In this attack, $\adversaryClient$ adds manipulated data $D^\adversaryClient$ to the training datasets of compromised clients $i$ by flipping data labels, e.g., by changing the labels of a street sign database so that pictures showing a 30~km/h speed limit are labeled as 80~km/h \cite{shen}, or, by adding triggers into data samples (e.g., a specific pixel pattern added to images~\cite{xie2020dba}) in combination with label flipping.}{}
We denote the fraction of injected poisoned data $ D^\adversaryClient_i$ in the overall poisoned training \mbox{dataset $ D'_i $ of client $i$ as \emph{Poisoned Data Rate (PDR)}, i.e.,} $PDR_i = \frac{\vert D^\adversaryClient_i \vert}{\vert D'_i \vert}$.	

\vspace{0.2em}
\noindent \textbf{Model Poisoning.}
\label{sec:model-poisoning}
This attack technique requires that $\adversaryClient$ can fully control a number of clients. $\adversaryClient$ poisons the training datasets of these clients and manipulates how they execute the training process by modifying parameters and scaling the resulting model update to maximize the attack impact while evading the aggregator's anomaly detector \cite{bagdasaryan, wang2020attack}. This is done by (1) scaling up the weights of malicious model updates to maximize attack impact (e.g., \modelreplacement attack \cite{bagdasaryan}, or, \emph{projected gradient descent (PGD) attack with model replacement} \cite{wang2020attack}), or, scaling down model updates to make them harder to detect (e.g., \trainandscale \cite{bagdasaryan} ) and (2) constraining the training process itself to minimize the deviation of malicious models from benign models to evade anomaly detection (e.g., \constrainandscale attack \cite{bagdasaryan}).  

\vspace{-1em}
\subsection{Adversary Goals and Capabilities}
\label{sec:adversary-model}
\vspace{-0.5em}
The goals of the adversary are two-fold:

\gAccuracy: The adversary $\adversaryClient$ aims to manipulate the global model $ G $ so that the modified model $ G' $ provides incorrect predictions $ f(G',x) = c' \neq f(G,x) $ for any inputs $ x \in I_\adversaryClient$, where $I_\adversaryClient$ is the so-called \emph{trigger set} consisting of specific attacker-chosen inputs and $c'$ denotes the incorrect prediction chosen by the adversary.

\gStealthiness: To make the poisoned model $G'$ hard to detect by aggregator $A$, it should closely mimic the behavior of $G$ on all other inputs not in $I_\adversaryClient$, i.e.: 
\vspace{-0.6em}
\begin{equation}
\label{eq:adv-goal}
f(G',x)=\left\{\begin{matrix}
c' \neq f(G,x)  & \forall x \in I_\adversaryClient \\ 
f(G,x) &  \forall x \notin I_\adversaryClient
\end{matrix}\right.
\vspace{-0.6em}
\end{equation}
	
Additionally, to make poisoned models as indistinguishable as possible from benign models, the distance (e.g., euclidean) between a poisoned model $W'$ and a benign model $W$ must be smaller than a threshold $\eta$ denoting the distinction capability of the anomaly detector of aggregator $A$, i.e., 
    $\dist(W, W') <\eta $.   
\revisionAdded{The adversary can estimate this distance by comparing the local malicious model to the global model or to a local model trained on benign data.\\}{}

\vspace{-0.6em}
\noindent\textbf{Adversarial Capabilities.} 
In this paper, we make no specific assumptions about the adversary's behavior.
We assume that the adversary $\adversaryClient$ has full control over $k < \frac{n}{2}$ clients and their training data, processes, and parameters~\cite{bagdasaryan,xie2020dba}. We denote the fraction of compromised clients as Poisoned Model Rate $\pmr=\frac{k}{n}$. Furthermore, $\adversaryClient$ has full knowledge of the aggregator's operations, including potentially applied backdoor defenses. 
However, $\adversaryClient$ has no control over any processes executed at the aggregator nor over the honest clients.

\vspace{-1em}
\subsection{Preliminaries}  
\label{sec:prelim}
\vspace{-0.3em}

\revisionAdded{\paragraph{\hdbscan~\cite{campello2013HDBSCAN}} 
 is a density-based clustering algorithm that uses the distance of data points in $n$-dimensional space to group data points that are located near each other together into a cluster. Hereby the number of clusters is determined dynamically. Data points that do not fit to any cluster are considered outliers. However, while HDBSCAN's predecessor \dbscan~\cite{ester1996density}  uses a predefined maximal distance to determine whether two points belong to the same cluster, \hdbscan determines this maximal distance for each cluster independently, based on the density of points. Thus, in \hdbscan, neither the maximal distance nor the total number of clusters need to be predefined.}

\vspace{-1.3em}
\revisionAdded{\paragraph{Differential Privacy (DP).} DP is a privacy technique that aims to ensure that the outputs do not reveal
individual data records of participants. DP is formally defined as follows:}{}

\vspace{-0.25em}

\begin{definition}[\revisionAdded{$(\epsilon,\delta)$-differential privacy})]
\label{def:DP}
\revisionAdded{A randomized algorithm $\mathcal{M}$ is
$(\epsilon,\delta)$-differentially private if for any datasets $D_1$ and $D_2$ that differ on a single element, and any subset of outputs $\mathcal{S} \in Range(\mathcal{M})$, the following inequality holds: }

    \begin{equation*}
        Pr[\mathcal{M}(D_1) \in \mathcal{S}] \leq e^\epsilon \cdot Pr[\mathcal{M}(D_2) \in \mathcal{S}] + \delta.
    \end{equation*}
\end{definition}

\noindent \revisionAdded{Here, $\epsilon$ denotes the privacy bound and $\delta$ denotes the probability of breaking this bound~\cite{dwork2014algorithmic}. Smaller values of $\epsilon$ and $\delta$ indicate stronger privacy.
A commonly used approach to enforce differential privacy is adding random Gaussian noise ${N}(0,\sigma^2)$ to the output of the algorithm~\cite{abadi2016ccsDifferential,dwork2014algorithmic}. 
}{}

\vspace{-0.5em}
\section{Problem Setting and Objectives}
\label{sec:backdoor_types}

\vspace{-0.9em}

\begin{figure}[t]
	\centering
	\setlength{\belowcaptionskip}{-8pt}
	\scalebox{0.7}{\begin{tikzpicture}

\newcommand{\rightMost}{8};
\newcommand{\leftMost}{0};

\draw[->, thick](0.1,-3) -- (3.8,-3);
\node[align=left] at (4.2,-3.1) {$\overrightarrow{G_{t-1}}$};

\draw[->, thick, red](0.1,-3) -- (2,- 0.2);
\node[align=left] at (2.2,0) {$\overrightarrow{W'_1}$};

\draw[->, thick, blue](0.1,-3) -- (4.5,-1);

\draw[->, thick, red](0.1,-3) -- (6.6,-1.1);
\node[align=left] at (7,-1) {$\overrightarrow{W'_2}$};

\draw[->, thick, red](0.1,-3) -- (4.4,-2.3);
\node[align=left] at (4.6,-2) {$\overrightarrow{W'_3}$};

\draw[->, thick, blue](0.1,-3) -- (4.3,-2.8);

\draw[->, thick, orange](3.8,-3) -- (2.04,- 0.2);
\node[align=left] at (3.5,-0.5) {$\overrightarrow{W'_1} - \overrightarrow{G_{t-1}}$};

\draw[->, thick, orange](3.8,-3) -- (6.65,-1.09);
\node[align=left] at (6.5,-2) {$\overrightarrow{W'_2} - \overrightarrow{G_{t-1}}$};

\draw[->, thick, orange](3.8,-3) -- (4.45,-2.3);
\node[align=left] at (5.6,-2.7) {$\overrightarrow{W'_3} - \overrightarrow{G_{t-1}}$};

\node[draw, align=left] at (6.55, 0.3) { -  Benign models\\  -  Backdoored models\\ -  Deviations of \\  Backdoored models};
\draw [fill=blue] (5,1) rectangle (5.2,0.8);
\draw [fill=red] (5, 0.6) rectangle (5.2,0.4);
\draw [fill=orange] (5,0.2) rectangle (5.2,0);

\end{tikzpicture}













}
	\vspace{-1.em}
 	\caption{Weight vectors of benign and backdoored models.}
 	\label{fig:backdoor_types}
 	\label{fig:model-vectors}
 	\label{fig:vector-distance}
 	\vspace{-0.3cm}
\end{figure}
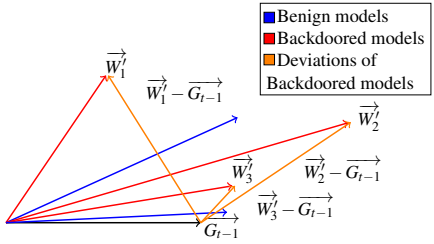

\noindent\textbf{Backdoor Characterization.} \revisionAdded{Following common practice in FL-related papers (e.g., \cite{bagdasaryan, fung, cao2020fltrust}), we represent Neural Networks (NNs) using their weight vectors, in which the extraction of weights is done identically for all models by flattening/serializing the weight/bias matrices in a predetermined order.}{} Figure~\ref{fig:backdoor_types} shows an abstract two-dimensional representation of the weight vectors of local models compared to the global model $G_{t-1}$ of the preceding aggregation round. \minorRevision{Each model $W_i$ can be characterized with two factors: \textit{direction (angle)} and \textit{magnitude (length)} of its weight vector $(w^1, w^2, \ldots, w^p)$. The angle between two updates $W_i$ and $W_j$ can be measured, e.g., by using the cosine distance metric $c_{ij}$ as defined in (\ref{eq:cosine-distance}) while their magnitude difference is measured by the \lnorm $e_{ij}$ as defined in (\ref{eq:l2norm-distance}).
\vspace{-0.6em}
\begin{equation}
 		\label{eq:cosine-distance}
 			c_{ij} = 1- \frac{W_iW_j}{\left \| W_i\right \| \left \| W_j\right \|} 
 		 	   		= 1- \frac{\sum_{k=1}^{p}{w_i^k}w_j^k}
 					{\sqrt{\sum_{k=1}^{p}({w_i^k })^2}\;\sqrt{\sum_{k=1}^{p}({w_j^k})^2}}
 	\end{equation}

\begin{equation}
		\label{eq:l2norm-distance}
 			e_{ij} =\left \| W_i - W_j \right \|			
 					= \sqrt{\sum_{k=1}^p{{(w_i^k - w_j^k)}^2}}
 \end{equation}
 	}{}
Benign and backdoored local models are shown in blue and red colors and are labeled with $W_i$ or $W_i'$, respectively. Note that the benign models are typically not identical due to the potentially partially \noniid nature of their training data.

The impact of the adversarial goal (injection of a backdoor) causes a deviation in the model parameters that manifests itself as a difference in the direction and/or magnitude of the backdoored model's weight vector in comparison to benign models, e.g., the deviations among local models and to the global model $G_{t-1}$ of the previous aggregation round. Since the adversary has full control over the training process of compromised clients, he can fully control these distances, e.g., by changing the direction (in the case of $W'_1$) or magnitude (in the case of $W'_2$) of the backdoored models' weight vectors. 

Figure~\ref{fig:backdoor_types} also shows three kinds of backdoored models resulting from different types of backdoor attacks. The first type $W'_1$ has a similar weight vector, but a \textit{large angular deviation} from the majority of local models and the global model. This is because such models are trained to obtain high accuracy on the backdoor task, which can be achieved by using a large poisoned data rate ($PDR$) or a large number of local training epochs (cf. Distributed Backdoor Attack (DBA)~\cite{xie2020dba}). 
The second backdoor type $W'_2$ has a small angular deviation but a \textit{large magnitude} to amplify the impact of the attack. Such models can be crafted by the adversary by \emph{scaling up} the model weights to boost its effect on the global model (cf. \Modelreplacement attack in \cite{bagdasaryan}). The third backdoor type $W'_3$ has a similar weight vector as benign models, the angular difference and the magnitude are not substantially different compared to benign models and, thus less distinguishable from benign models. Such \textit{stealthy} backdoored models can be crafted by the adversary by carefully constraining the training process or scaling down the poisoned model's weights (cf. \Constrainandscale attack \cite{bagdasaryan} or FLIoT attack \cite{nguyen2020diss}).

\noindent {\bf Defense Objectives.}  \label{sec:defense-goals}
A generic defense that can effectively mitigate backdoor attacks in the FL setting needs to fulfill the following objectives: (i) \textit{Effectiveness}: To prevent the adversary from achieving its attack goals, the impact of backdoored model updates must be eliminated so that the aggregated global model does not demonstrate backdoor behavior. (ii) \textit{Performance}: Benign performance of the global model must be preserved to maintain its utility. (iii) \textit{Independence from data distributions and attack strategies}: The defense method must be applicable to generic adversary models, i.e., it must not require prior knowledge about the backdoor attack method, or make assumptions about specific data distributions of local clients, e.g., whether the data are \iid or \noniid.

\vspace{-0.5em}
\section{\ourname Overview and Design}
\label{sec:idea}
\vspace{-1em}
We present the high-level idea of \ourname and the associated design challenges to fulfill the objectives identified in~\sect \ref{sec:backdoor_types}.

\vspace{-1.3em}
\subsection{High-level Idea}  \label{sec:high-level}
\vspace{-0.5em}

\noindent \textbf{Motivation.} Earlier works (e.g., Sun~\etal~\cite{sun2019can})  \revisionChanged{use differential privacy-inspired noising of the aggregated model for eliminating backdoors. They determine the sufficient amount of noise to be used empirically. In the FL setting this is, however, challenging, as one cannot in general assume the aggregator to have access to training data, in particular to poisoned datasets. What is therefore needed is a generic method for determining how much noise is sufficient to remove backdoors effectively. On the other hand, the more noise is injected into the model, the more its benign performance will be impacted.}{}

\begin{figure}[t]
    \vspace{-0.1cm}
    \setlength{\belowcaptionskip}{-15pt}
	\centering{
	   \subfloat[Applying clustering and clipping.]{
	   \scalebox{0.7}{
	        \begin{tikzpicture}

\newcommand{\rightMost}{8};
\newcommand{\leftMost}{0};


\draw[->, thick](0.1,-3) -- (3.8,-3);
\node[align=left] at (4.2,-3.1) {$\overrightarrow{G_{t-1}}$};

\draw[->, thick, red](0.1,-3) -- (1.7, 0);
\node[align=left] at (1.9,0.2) {$\overrightarrow{W'_1}$};

\draw[black](0.75,-0.7) -- (1.75,-1.1);
\draw[black](0.75,-1.1) -- (1.75,-0.7);

\draw[->, thick, blue](0.1,-3) -- (4.8,-1);

\draw[->, thick, red](0.1,-3) -- (6.6,-1.2);
\node[align=left] at (6.9,-1.2) {$\overrightarrow{W'_2}$};

\draw[->, thick, red](0.1,-3) -- (4,-2.3);
\node[align=left] at (4.25,-2.3) {$\overrightarrow{W'_3}$};

\draw[->, thick, blue](0.1,-3) -- (4.1,-2.8);


\fill[cyan] (0.05,-0.094) circle (0.2cm);
\node[align=left] at (0.05,-0.094) {\textbf{\textcolor{white}{1}}};
\node[align=left] at (0.9,-0.1) {Filtering};

\fill[cyan] (5.15,-0.64) circle (0.2cm);
\node[align=left] at (5.15,-0.64) {\textbf{\textcolor{white}{2}}};
\node[align=left] at (6.0,-0.7) {Clipping};
\draw [->, dashed, thick] (6.6,-1.1) to [out=150,in=30] (4.45,-1.7);
\draw [->, dashed, thick] (4.8,-0.9) to [out=150,in=60] (4.25,-1.15);


\node[align=left] at (3.2,-0.5) {Clipping bound $S$};
\draw[thick, dashed] (4.65,-3.2) arc (0:32 :4.6cm);


\node[draw, align=left] at (6.55,0.2) { -  Benign models\\  -  Backdoored models};
\draw [fill=blue] (5,0.5) rectangle (5.2,0.3);
\draw [fill=red] (5,0.1) rectangle (5.2,-0.1);

\end{tikzpicture}}
	        \label{fig:our-idea}
	    }
	    \vspace{-0.25cm}\\
		\subfloat[Applying noising]{
	   \scalebox{0.7}{
			\begin{tikzpicture}

\draw[->](0,-2) -- (1.8,-2);
\node[align=left] at (2.2, -2.1) {\textbf{$\overrightarrow{G_{t-1}}$}};

\draw[->, blue](0,-2) -- (2,-0.3);
\draw[->, red](0,-2) -- (2.3,-0.8);
\draw[->, red](0,-2) -- (2.3,-1.4);
\draw[->, blue](0,-2) -- (2.3,-1.8);
\node[align=left] at (2.2,-0.1) {\textbf{$S$}};
\draw[thick, dashed] (2.6,-2) arc (0:45:2.6cm);
\draw[->, ultra thick](0,-2) -- (2.3,-1.1);
\node[align=left] at (2.7, -1.1) {\textbf{$\overrightarrow{G_t}$}};



\draw[->](4.2,-1.8) -- (6,-1.8);
\node[align=left] at (6.4, -1.9) {\textbf{$\overrightarrow{G_{t-1}}$}};

\draw[->, ultra thick](4.2,-1.8) -- (6.5,-0.9);
\node[align=left] at (6.65, -1.15) {\textbf{$\overrightarrow{G_t}$}};

\fill[cyan] (6,0) circle (0.2cm);
\node[align=left] at (6,0) {\textbf{\textcolor{white}{3}}};
\node[align=left] at (6.8,0) {Noising};
\draw[green] (6.5,-0.9) circle (0.6cm);
\draw[green] (6.5,-0.9) circle (0.01cm);
\draw[green, thick](6.5,-0.9) -- (6.5,-0.3);
\node[align=left] at (6.3,-0.65) {$\sigma$};

\node[draw, align=left] at (4.2,-0.1) { $\sigma$: Noise level};

\end{tikzpicture}}
			\label{fig:after-defenses}
		}
	\vspace{-0.7em}
	\caption{High-level idea of \ourname defense. \label{fig:idea}
	}}\vspace{-0.05cm}
\end{figure}
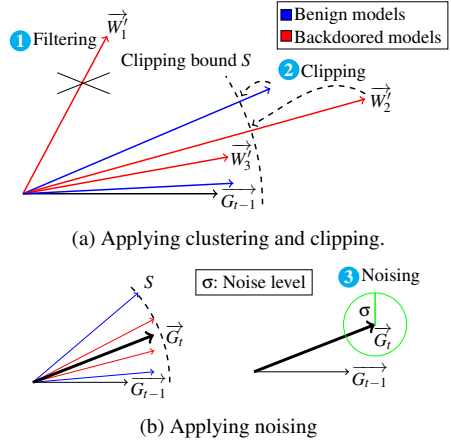

\vspace{0.15em}
\noindent \textbf{\ourname Overview.} \revisionChanged{\ourname estimates the noise level required for backdoor removal in the FL setting without extensive empirical evaluation and having access to training data (this noise bound is formally proven in~\sect\ref{sec:security-analysis}).}{}
In addition, to effectively limit the amount of required noise, \revisionChanged{\ourname uses}{} a novel clustering-based approach to identify and remove \revisionChanged{adversarial}{} model updates with high impact and \revisionChanged{applies}{} a dynamic weight-clipping approach to limit the impact of models that the adversary has scaled up to boost their performance. \revisionAdded{As discussed in \sect\ref{sec:backdoor_types}, one cannot guarantee that all backdoored models can be detected since the adversary can fully control both the angular and magnitude deviation to make the models arbitrarily hard to detect. Our clustering approach therefore aims to remove models with high attack impact (having larger angular deviation) rather than all malicious models.}{}
Fig.~\ref{fig:idea} illustrates the high-level idea of \ourname consisting of the above three components: filtering, clipping, and noising.
We emphasize, however, that each of these components needs to be applied with great care, since, a na\"ive combination of noising with clustering and clipping leads to poor results as it easily fails to mitigate the backdoor and/or deteriorates the benign performance of the model, as we show in \sect\ref{app:naive-combination}.
 We detail the design of each component and its use in the \ourname defense approach in~\sect \ref{sec:defense}.

\vspace{-1em}
\subsection{Design Challenges} \label{sec:design-challenges}
\vspace{-0.4em}

To realize the high-level idea presented above, we need to solve the following technical challenges.

\noindent\textbf{$C_1$- Filtering out backdoored models with large angular deviations in dynamic scenarios.} 
As discussed in \sect\ref{sec:backdoor_types}, the weight vector of a well-trained backdoored model, $W'$, has a \textit{higher angular difference} in comparison to weight vectors of benign models $W$. 
\revisionChanged{\ourname deploys a clustering approach to \textit{identify} such poisoned models and \textit{remove} them from FL aggregation (detailed in~\sect\ref{sec:model-C}). 
The effect of clustering-based filtering is shown in Fig.~\ref{fig:our-idea} where model $W'_1$ is removed from the aggregated model as it does not align with the directions of benign models.}{}
In contrast to existing clustering-based defenses, we need an approach that can also work in a \emph{dynamic attack setting}, i.e., the number of injected backdoors is unknown and may vary between training rounds. To this end, we make a key observation: clustering approaches using a fixed number of clusters $n_{cluster}$ for identifying malicious models are inherently vulnerable to attacks with varying numbers of backdoors\footnote{We consider two backdoors to be independent if they use different triggers.} $n_{backdoor}$. This is because the adversary can likely cause at least one backdoor model to be clustered together with benign models due to the pigeonhole principle by simultaneously injecting $n_{backdoor} \geq n_{cluster}$ backdoors. 
We seek to solve this challenge by employing a clustering solution that dynamically determines the clusters for model updates, thereby allowing it to adapt to dynamic attacks.

\noindent\textbf{$C_2$-Limiting the impact of scaled-up backdoors.} 
To limit the impact of backdoored models that the adversary artificially scales up to boost the attack (e.g., $W'_2$ in Fig.~\ref{fig:model-vectors}), the weight vectors of models with high magnitudes can be clipped~\cite{sun2019can}. 
The effect of clipping is shown in Fig.~\ref{fig:our-idea} where the weight vectors of all models with a magnitude beyond the clipping bound $S$ (in particular, backdoored model $W'_2$) are clipped to $S$ by scaling down the weight vectors. 
The resulting clipped weight vectors are shown on the left side of Fig.~\ref{fig:after-defenses}.
The challenge here is how to select a proper clipping bound without empirically evaluating its impact on the training datasets (which are not available in the FL setting). 
If the applied clipping bound is too large, an adversary can boost its model $W'$ by scaling its weights up to the clipping bound, thereby maximizing the backdoor impact on the aggregated global model $G$. However, if the applied clipping bound is too small, a large fraction of benign model updates $W$ will be clipped, thereby leading to performance deterioration of the aggregated global model $G$ on the main task.
We tackle this challenge in \sect\ref{sec:clipping-noising}, where we show how to select a clipping bound that can not be influenced by the adversary and that effectively limits the impact of scaled-up backdoored models.

\vspace{0.1em}
\noindent\textbf{$C_3$-Selecting suitable noise level for backdoor elimination.} 
As mentioned in \sect\ref{sec:high-level}, \ourname uses model noising that applies Gaussian noise with noise level $\sigma$ to mitigate the adversarial impact of backdoored models (e.g., $W'_3$ in Fig.~\ref{fig:vector-distance}).
Similar to the clipping bound, however, also here the noise level $\sigma$ must be carefully selected, as it has a direct impact on the effectiveness of the defense and the model's benign performance. If it is too low, the aggregated model might retain backdoor behavior after model noising, rendering the defense ineffective, while excessive noise  
will degrade the utility of the aggregated model.
To address this challenge, we develop an approach for reliably estimating a sufficient but minimal bound for the applied noise in~\sect\ref{sec:security-analysis}.

\vspace{-1.2em}
\subsection{\ourname Design}
\label{sec:defense}

\begin{figure}[t]
	\centering
	\includegraphics[width=0.75\columnwidth]{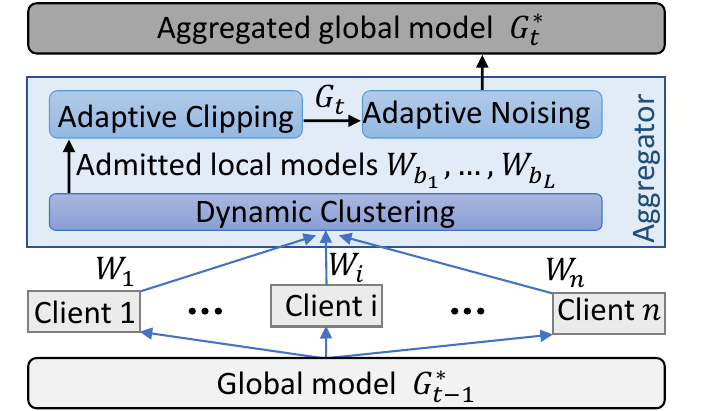}
	\vspace{-0.5em}
	\caption{Illustration of \ourname's workflow in round $t$.}
	\label{fig:defense-process}
\end{figure}

\vspace{-0.8em}
As discussed in \sect\ref{sec:high-level}, our defense consists of three main components: filtering, clipping, and noising. Figure \ref{fig:defense-process} shows these components and the workflow of \ourname during training round $t$. Algorithm~\ref{alg:ouralg} outlines the procedure of \ourname. 
In the rest of this section, we detail the design of these components to resolve the challenges in \sect\ref{sec:design-challenges}.

\vspace{-0.9em}
\setlength{\textfloatsep}{0.1em}
\begin{algorithm}[ht!]
\small
	\caption{\ourname}
	\label{alg:ouralg} 
	\begin{algorithmic}[1]
		\State \textbf{Input: $n$,  $G_0$, $T$} \Comment{ $n$ is the number of clients, $G_0$ is the initial global model, $T$ is the number of training iterations}
		\State \textbf{Output: $G^*_T$} \Comment{$G^*_T$ is the updated global model after $T$ iterations}
		\For{each training iteration $t$ in $[1, T]$}
		\For {each client $i$ in $[1, n]$}
		    \State $W_i \gets$ \Call{ClientUpdate}{$G^*_{t-1}$}  \Comment{The aggregator sends $G^*_{t-1}$ to Client $i$ who trains $G^*_{t-1}$ using its data $D_i$ locally to achieve local modal $W_i$ and sends $W_i$ back to the aggregator.}
		\EndFor		
		
		\State $(c_{11}, \ldots, c_{nn}) \gets$ \Call{CosineDistance}{$W_1,\ldots, W_n$} \Comment{$\forall i,j \in (1, \ldots, n)$, $c_{ij}$ is the cosine distance between $W_i$ and $W_j$} 	
		
		\State $(b_1, \ldots, b_L) \gets$ \Call{Clustering}{$c_{11}, \ldots, c_{nn}$}  \Comment{\mbox{$L$ is the} number of admitted models, $b_l$ is the index of the $l^{th}$ model}
		
		\State $(e_1, \ldots, e_n) \gets$ \Call{EuclideanDistances}{$G^*_{t-1}, (W_1,\ldots, W_n)$} 
		\Comment{ $e_i$ is the Euclidean distance between $G^*_{t-1}$ and $W_i$} 		
		\State $S_t \gets \Call{Median}{e_1, \ldots, e_n}$  \Comment{$S_t$ is the adaptive clipping bound at round $t$}
		\For {each client $l$ in $[1, L]$}
		\State $W^c_{b_l} \gets G_{t-1} + (W_{b_l} - G_{t-1}) \cdot \Call{Min}{1, \gamma }$
		\Comment{Where $\gamma$ ($= S_t/e_{b_l}$) is the clipping parameter, $W^c_{b_l}$ is the admitted model after clipped by the adaptive clipping bound $S_t$}
		\EndFor
		
		\State $G_{t} \gets \sum_{l=1}^L W^c_{b_l}/L$ \Comment{Aggregating, $G_{t}$ is the plain global model before adding noise}
		
		\State $\sigma \gets \lambda \cdot S_t$ where $ \lambda = \frac{1}{\epsilon} \cdot \sqrt{2ln \frac{1.25}{\delta}}$
		\Comment{Adaptive noising level}
		
		\State $G^*_t \gets G_t + N(0, \sigma^2)$ \Comment{Adaptive noising}
		\EndFor
	\end{algorithmic}
\end{algorithm}

\vspace{-0.6em}
\subsubsection{Dynamic \ModelFiltering}
\label{sec:model-C}
\begin{figure}[t]
\vspace{-0.4cm}
	\centering{
		    \subfloat[Ground truth]{
		    
	        \scalebox{0.75}{
	            \begin{tikzpicture}


\draw[->, blue](1.5,-2) -- (3.3,-1.4);
\draw[->, blue](1.5,-2) -- (3.3,-1.5);
\draw[->, blue](1.5,-2) -- (3.3,-1.6);
\draw[->, blue](1.5,-2) -- (3.3,-1.7);
\draw[->, blue](1.5,-2) -- (3.3,-1.8);
\draw[blue, dashed, thick]  plot[smooth, tension=.7] coordinates {(1.5,-2) (2.5, -1.5) (3.4,-1.3) (3.5,-1.6) (3.4, -1.8) (2.5,-2) (1.5,-2)};

\node[align=left] at (3.1, -2.3) {\textcolor{blue}{Benign}};

\draw[->, red](1.5,-2) -- (3,-0.8);
\draw[->, red](1.5,-2) -- (3,-1);

\draw[red, dashed, thick]  plot[smooth, tension=.7] coordinates {(1.5,-2) (2.5, -1) (3,-0.7) (3.2,-0.9) (3.1, -1.1) (2.5,-1.5) (1.5,-2)};

\node[align=left] at (2.5, -0.4) {\textcolor{red}{Backdoored}};

\draw[->,  orange](1.5,-2) -- (0.2,-1.1);
\draw[->, orange](1.5,-2) -- (0.2,-1.3);

\draw[orange, dashed, thick]  plot[smooth, tension=.7] coordinates {(1.5,-2) (0.8, -1.2) (0.1,-1) (0,-1.2) (0.1, -1.4) (0.8,-1.8) (1.5,-2)};

\node[align=left] at (0.4, -0.6) {\textcolor{orange}{Random}};

\end{tikzpicture}
	            \label{fig:cluster-example:groundtruth}
	        }
	    } 
	       	\subfloat[K-means]{
	        \scalebox{0.75}{
			    \begin{tikzpicture}


\draw[->, blue](1.5,-2) -- (3.3,-1.4);
\draw[->, blue](1.5,-2) -- (3.3,-1.5);
\draw[->, blue](1.5,-2) -- (3.3,-1.6);
\draw[->, blue](1.5,-2) -- (3.3,-1.7);
\draw[->, blue](1.5,-2) -- (3.3,-1.8);
\draw[blue, dashed, thick]  plot[smooth, tension=.7] coordinates {(1.5,-2) (2.5, -1) (3,-0.7) (3.2,-0.9) (3.4,-1.2) (3.5,-1.6) (3.4, -1.8) (2.5,-2) (1.5,-2)};

\node[align=left] at (3, -2.3) {\textcolor{blue}{Accepted}};

\draw[->, red](1.5,-2) -- (3,-0.8);
\draw[->, red](1.5,-2) -- (3,-1);



\draw[->,  orange](1.5,-2) -- (0.2,-1.1);
\draw[->, orange](1.5,-2) -- (0.2,-1.3);

\draw[red, dashed, thick]  plot[smooth, tension=.7] coordinates {(1.5,-2) (0.8, -1.2) (0.1,-1) (0,-1.2) (0.1, -1.4) (0.8,-1.8) (1.5,-2)};

\node[align=left] at (0.4, -0.6) {\textcolor{red}{Rejected}};


\end{tikzpicture}
			    \label{fig:cluster-example:kmean}
	        } 
		} \\
		\vspace{-.3cm}
	    \subfloat[\hdbscan]{
	        \scalebox{0.75}{
	            \begin{tikzpicture}


\draw[->, blue](1.5,-2) -- (3.3,-1.4);
\draw[->, blue](1.5,-2) -- (3.3,-1.5);
\draw[->, blue](1.5,-2) -- (3.3,-1.6);
\draw[->, blue](1.5,-2) -- (3.3,-1.7);
\draw[->, blue](1.5,-2) -- (3.3,-1.8);
\draw[black, dashed, thick]  plot[smooth, tension=.7] coordinates {(1.5,-2) (2.5, -1.5) (3.4,-1.3) (3.5,-1.6) (3.4, -1.8) (2.5,-2) (1.5,-2)};

\node[align=left] at (3, -2.3) {Cluster A};

\draw[->, red](1.5,-2) -- (3,-0.8);
\draw[->, red](1.5,-2) -- (3,-1);

\draw[black, dashed, thick]  plot[smooth, tension=.7] coordinates {(1.5,-2) (2.5, -1) (3,-0.7) (3.2,-0.9) (3.1, -1.1) (2.5,-1.5) (1.5,-2)};

\node[align=left] at (2.5, -0.4) {Cluster B};

\draw[->,  orange](1.5,-2) -- (0.2,-1.1);
\draw[->, orange](1.5,-2) -- (0.2,-1.3);

\draw[black, dashed, thick]  plot[smooth, tension=.7] coordinates {(1.5,-2) (0.8, -1.2) (0.1,-1) (0,-1.2) (0.1, -1.4) (0.8,-1.8) (1.5,-2)};

\node[align=left] at (0.4, -0.6) {Cluster C};

\end{tikzpicture}
	            \label{fig:cluster-example:hdbscan}
	        } 
	    }
	    \subfloat[\ourname]{
	        \scalebox{0.75}{
	            \begin{tikzpicture}


\draw[->, blue](1.5,-2) -- (3.3,-1.4);
\draw[->, blue](1.5,-2) -- (3.3,-1.5);
\draw[->, blue](1.5,-2) -- (3.3,-1.6);
\draw[->, blue](1.5,-2) -- (3.3,-1.7);
\draw[->, blue](1.5,-2) -- (3.3,-1.8);
\draw[blue, dashed, thick]  plot[smooth, tension=.7] coordinates {(1.5,-2) (2.5, -1.5) (3.4,-1.3) (3.5,-1.6) (3.4, -1.8) (2.5,-2) (1.5,-2)};

\node[align=left] at (3, -2.3) {\textcolor{blue}{Accepted}};

\draw[->, red](1.5,-2) -- (3,-0.8);
\draw[->, red](1.5,-2) -- (3,-1);


\node[align=left] at (1.5, -0.6) {\textcolor{red}{Rejected (Outliers)}};

\draw[->,  red](1.5,-2) -- (0.2,-1.1);
\draw[->, red](1.5,-2) -- (0.2,-1.3);



\end{tikzpicture}
	            \label{fig:cluster-example:flguard}
	        } 
	    } 

	}
	\vspace{-1em}
	\caption{\small{Comparison of clustering quality for \protect\subref{fig:cluster-example:groundtruth} ground truth, \protect\subref{fig:cluster-example:kmean} using \kmeans with 2 clusters as in Auror~\cite{shen}, \protect\subref{fig:cluster-example:hdbscan} straightforward applied \hdbscan and \protect\subref{fig:cluster-example:flguard} our approach as in \ourname.} 
	}
	\label{fig:cluster-example}
\end{figure}
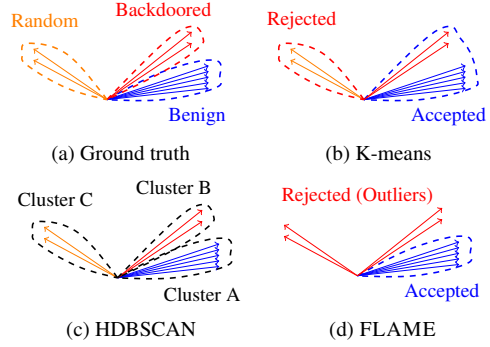

\vspace{-0.6em}
The \ModelFiltering component of \ourname utilizes a \textit{dynamic clustering} technique based on \hdbscan~\cite{campello2013HDBSCAN} that identifies poisoned models with high angular deviations from the majority of updates (e.g., $W'_1$ in Fig.~\ref{fig:our-idea}). 
Existing clustering-based defenses~\cite{blanchard,shen} identify potentially malicious model updates by clustering them into two groups where the smaller group is always considered malicious and thus removed. However, if no malicious models are present in the aggregation, this approach may lead to many models being incorrectly removed and thus a reduced accuracy of the aggregated model. These approaches also do not protect against attacks in which adversary $\adversaryClient$ simultaneously injects multiple backdoors by using different groups of clients to inject different backdoors. If the number of clusters is fixed, there is the risk that poisoned and benign models end up in the same cluster, in particular, if models with  different backdoors differ significantly. Consequently, existing model clustering methods do not adequately address challenge $C_1$ (\sect\ref{sec:design-challenges}).
Fig.~\ref{fig:cluster-example} shows the behavior of different clustering methods on a set of model updates' weight vectors.
Fig.~\ref{fig:cluster-example:groundtruth} shows the ground truth of an attack scenario where $\adversaryClient$ uses two groups of clients: one group is used to inject a backdoor, whereas the other group provides random models with the goal of fooling clustering-based defenses. Fig.~\ref{fig:cluster-example:kmean} shows how in this setting, \kmeans (as used in Auror~\cite{shen}) fails to successfully separate benign and poisoned models as all poisoned models end up in the same cluster with the benign models.

\vspace{-0.2em}
To overcome the limitations of existing defenses, we design our clustering solution and ensure that:
(i) it is able to handle dynamic attack scenarios where multiple backdoors are injected simultaneously, and (ii) it minimizes false positives of poisoned model identification. 
In contrast to existing approaches that try to place poisoned models into one cluster, our approach considers each poisoned model individually as an outlier, so that it can gracefully handle multiple simultaneous backdoors and thus address challenge $C_1$.

\vspace{-0.2em}
\ourname uses pairwise cosine distances to measure the angular differences between all model updates and applies the \hdbscan clustering algorithm~\cite{campello2013HDBSCAN}. The advantage here is that cosine distances are not affected even if the adversary scales up model updates to boost their impact as this does not change the angle between the updates' weight vectors. 
\revisionChanged{Since the \hdbscan algorithm clusters the models based on their {density of the cosine distance distribution} and \emph{dynamically determines the required number of clusters}, we leverage it for our dynamic clustering approach. We describe \hdbscan and how we apply it in detail in \sect\ref{sec:HDBSCAN}. 

In particular, \hdbscan labels models as outliers if they do not fit into any cluster. This allows \ourname to effectively handle multiple poisoned models with different backdoors by labeling them as outliers.}{}
To realize this, we set the minimum cluster size 
to be at least $50\%$ of the clients, i.e., $\frac{n}{2}+1$, so that the resulting cluster will contain the majority of updates (which we assume to be benign, cf. \sect\ref{sec:adversary-model}). 
All remaining (potentially poisoned) models are marked as \textit{outliers}. This behavior is depicted in Fig.~\ref{fig:cluster-example:flguard} where all the models from Clusters B and C from Fig.~\ref{fig:cluster-example:hdbscan} are considered as outliers.  
Hence, to the best of our knowledge, our approach is the first FL backdoor defense that is able to gracefully handle also dynamic attacks in which the number of injected backdoors may vary.
The clustering step is shown in lines 6-7 of Alg.~\ref{alg:ouralg} where $L$ models are retained after clustering.

\vspace{-1.2em}
\subsubsection{Adaptive Clipping and Noising}
\label{sec:clipping-noising}
\vspace{-0.5em}

As discussed in \sect\ref{sec:design-challenges} (challenges $C_2$ and $C_3$), determining a proper clipping bound and noise level for model weight clipping and noising is not straightforward. 
We present our new approach for selecting an effective clipping bound and reliably estimating a sufficient noise level that can effectively eliminate backdoors while preserving the performance of the main task. 
Furthermore, our defense approach is resilient to adversaries that dynamically adapt their attacks.

\begin{figure}[t!]
	\centering
	\setlength{\belowcaptionskip}{-10pt}
	\includegraphics[width=0.6\columnwidth]{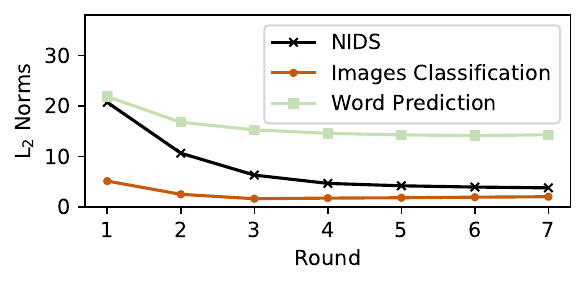}
	\vspace{-1.4em}
	\caption{\lnorm{s} of model updates depending on the number of training rounds for different datasets.}
	\vspace{1em}
  \label{fig:l2norm}
\end{figure}

\noindent\textbf{Adaptive Clipping.} 
Fig.~\ref{fig:l2norm} shows the variation of the average \lnorm{s} of model updates of benign clients in three different datasets (cf. \sect\ref{sec:dataset}) over subsequent training rounds.
We can observe that the \textit{\lnorm{s} of benign model updates become smaller in later training rounds}. 
To effectively remove backdoors while minimizing the impact on benign updates, the clipping bound $S$
needs to be dynamically adapted to this decreasing trend of the \lnorm. 
Recall that clipping is performed after clustering by scaling down model weights so that the \lnorm{} of the scaled model becomes smaller or equal to the clipping threshold. 
We describe how \ourname determines a proper scaling factor for each model update $W_i$ in $t^{th}$ training round as follows: Given the index set $(b_1, \ldots b_L)$ of the models admitted by the clustering method (line 7 of Alg.~\ref{alg:ouralg}), the aggregator first computes the clipping bound $S_t$ as the median of the \lnorm{}s of all $n$ model updates: $S_t = \Call{Median}{e_1, \ldots, e_n}$. It should be noted that for determining the clipping bound, the rejected models are also considered to ensure that even if benign models were filtered, the computed median $S_t$
is still determined based on benign values. However, after determining the clipping bound, only the admitted models $W_1,\ldots,W_L$ are considered for later processing. The scaling factor for the $l^{th}$ admitted model is computed as $\gamma = \frac{S_t}{e_{b_l}}$ where $e_{b_l}$ is the \lnorm{} of the model update $W_{b_l}$. Clipping scales down model updates as follows: $W^c_{b_l} = G_{t-1} + (W_{b_l} - G_{t-1}) \cdot \Call{Min}{1, \gamma }$ (detailed in line 8-11 of Alg.~\ref{alg:ouralg}) where the multiplication is computed coordinate-wise. \revisionAdded{It is worth noting that weighting contributions (i.e., adjusting scaling factor) based on client data sizes is insecure. As we point out in \sect\ref{sec:bg_fl}, the reported dataset sizes by clients cannot be trusted, i.e., the adversary can lie about their dataset sizes to maximize attack impact~\cite{wang2020attack}. Hence, we follow common practice in literature and weight the contributions of all clients equally regardless of their dataset size~\cite{blanchard, bagdasaryan, cao2020fltrust, xie2020dba}.}{}   
By using the median as the clipping bound $S_t$, we ensure that $S_t$ is always in the range of the $\lnorm$s between benign models and the global model since we assume that more than $50\%$ of clients are benign (cf.~\sect\ref{sec:adversary-model}). We evaluate the effectiveness of the clipping approach in~\sect\ref{app:eval-clipping}.

\noindent\textbf{Adaptive Noising.} 
\revisionAdded{It has been shown that by adding noise to a model's weights, the impact of outlier samples can be effectively mitigated~\cite{du2020iclr}.
Noise can also be added to poisoned samples (special cases of outliers) used in backdoor injection. The more noise is added to the model during the training process, the less responsive the model will be to the poisoned samples. Thus, increasing model robustness against backdoors. Eliminating backdoors utilizing noise addition is conceptually the same in a centralized or federated setting (e.g., \cite{bagdasaryan, du2020iclr}): In both cases, noise is added to the 
model weights to smooth out the effect of poisoned data (cf. Eq.~\ref{eq:local-to-global-noising}).}{}
\revisionAdded{The challenge is to determine as small a noise level as possible to eliminate backdoors and at the same time not deteriorate the benign performance of the model. As we discuss in detail in \sect\ref{sec:noise-proof}, the amount of noise is determined by estimating the sensitivity 
based on the differences (distances) among local models, which can be done without access to training data. We then}{} add Gaussian noise to the global model $G_t$ to yield a noised global model $G^*_t$ as follows: $G^*_t = G_t + N(0,\sigma^2)$, see Lines 13-14 of Alg. \ref{alg:ouralg} for more details. \revisionChanged{This ensures that backdoor contributions are effectively eliminated from the aggregated model.}{} \revisionAdded{In particular, we show in \sect\ref{sec:noise-proof} how the noise-based backdoor elimination technique can be transferred from a centralized to a federated setting by analysing the relationship between aggregated Gaussian noise applied to the global model and individual noising of each local model.}{}

\vspace{-0.5em}
\section{Security Analysis}
\label{sec:security-analysis}

\vspace{-0.5em}
\subsection{Noise Boundary Proof of \ourname}
\label{sec:noise-proof}
\vspace{-0.5em}
In this section, we provide a proof to corroborate that \ourname can neutralize backdoors in the FL setting by applying strategical noising with bound analysis on the noise level. 
We first formulate the noise boundary guarantee of \ourname in Theorem~\ref{thm:noise-scale}. Subsequently, we explain related parameters and prove how the noise level bound for $\sigma$ can be estimated. This is done by generalizing theoretical results from previous works~\cite{dwork2014algorithmic,du2020iclr} to the FL setting. 
Then, we show how the filtering and clipping component of \ourname helps to effectively reduce the noise level bound in Theorem~\ref{thm:reduce-noise}. 
\addressedComment{C1}
\minorRevision{We provide a formal proof for linear models and extend the proof to DNNs using empirical evaluation. This is because providing formal proof for DP-based backdoor security for DNN models is still an open research problem even for centralized settings.}{}

\vspace{-0.6em}
\begin{theorem}
\label{thm:noise-scale}
A $(\epsilon,\delta)$-differentially private model with parameters $G$ and clipping bound $S_t$
is backdoor-free if random Gaussian noise is added to the model parameters yielding a noised version $G^*$ of the model: $G^* \gets G + N(0, \sigma_G^2)$ where the noise scale $\sigma_G$ is determined by the clipping bound $S_t$ and a noise level factor $\lambda$: $\sigma_G \gets \lambda \cdot S_t$ and $ \lambda = \frac{1}{\epsilon} \cdot \sqrt{2ln \frac{1.25}{\delta}}$. 
\end{theorem}
\vspace{-0.4em}

We explore the key observation that an ML model with a sufficient level of differential privacy is \textit{backdoor-free}. With this new definition of backdoor-free models in the DP domain, the main challenge to defeat backdoors in the FL setting is to decide a proper noise scale for the global model without knowledge of the training datasets. Furthermore, we need to minimize the amount of noise added to the global model to preserve its performance on the main task. 
None of the prior DP-based FL backdoor defense techniques provide a solution to the noise determination problem~\cite{sun2019can}.     
For the first time, \ourname presents an approach to \textit{estimate} the proper noise scale that ensures the global model is backdoor-free. 
\minorRevision{The noise boundary proof in Theorem~\ref{thm:noise-scale} consists of two steps:}{}

\noindent \revisionChanged{\noindent \textbf{Step 1 (S1)}. By introducing the data hiding property of DP (Def.~\ref{def:DP}) and its \emph{implication as the theoretical guarantee for backdoor-free} models. }{}  
    We also discuss function sensitivity (Def.~\ref{def:func-sens}) which is an important factor for selection of the DP parameters $(\epsilon,\delta)$.  \\
\noindent \textbf{Step 2 (S2)}. We show how \ourname \textit{generalizes backdoor elimination from centralized setting to federated setting} with theoretical analysis of the noise boundary (\equ~\ref{eq:local-to-global-noising} and \ref{eq:global-noise}). \emph{\ourname is the first FL defense against backdoors that provides noise level proof with bounded backdoor effectiveness.
} 

\vspace{0.2em}
\noindent \textbf{(S1) DP foundations and re-interpretation as Backdoor-free.} 
\revisionChanged{As discussed in \sect\ref{sec:prelim}, by definition, DP makes the difference between data points indistinguishable.}{}  \ourname leverages this property of DP for backdoor elimination. In particular, we can consider $D_1$ and $D_2$ in Def.~\ref{def:DP} as the benign and backdoored dataset. The inequality of DP suggests that algorithm $\mathcal{M}$ has a high probability of producing the same outputs on the benign and the poisoned dataset, meaning that the backdoor is eliminated. 
The noise level $\sigma$ is determined based on the DP parameters $(\epsilon,\delta)$ and the \textit{sensitivity} of the function $f$ defined below: 
\vspace{-0.3em}
\begin{definition}[Sensitivity]
\label{def:func-sens}
Given the function $f:\mathcal{D} \rightarrow \mathbb{R}^d$ where $\mathcal{D}$ is the data domain and $d$ is the dimension of the function output, the sensitivity of the function $f$ is defined as:
\vspace{-0.5em}
\begin{equation}
    \label{eq:sens}
    \Delta = \underset{D_1,~D_2 ~\in ~\mathcal{D}}{max}~||f(D_1)-f(D_2)||_2,
    \vspace{-0.6em}
\end{equation}
where $D_1$ and $D_2$ differs on a single element $||D_1-D_2||_1=1$.
\end{definition}

\revisionAdded{As shown in Lemma 1 \cite{dwork2014algorithmic}, this definition can be extended to datasets differing by more than one element, i.e., can be generalized to the DP in the multiple-point-difference setting.}{}
 
\vspace{0.1em}
\noindent \textbf{(S2) Generalizing backdoor resilience from centralized to federated setting (\ourname).} 
In the centralized setting, the defender has access to the model to be protected, the benign dataset, and the outlier (backdoored) samples. As such, he can estimate the sensitivity $\Delta$ for $(\epsilon,\delta)$-DP. When applying Gaussian noise with the noise scale $\sigma=\frac{\Delta}{\epsilon} \sqrt{2ln \frac{1.25}{\delta}}$, the defender can enforce a lower bound on the prediction loss of the model on the backdoored samples for backdoor elimination [28]. However, this robustness rationale \textit{cannot} be directly transferred from the centralized setting to the FL setting since the defender in the federated scenario (i.e., aggregator) only has access to received model updates, but not the datasets to estimate the sensitivity $\Delta$ for the global model.

\revisionAdded{\ourname extends DP-based noising for backdoor elimination to the federated setting based on the following observation: if one can ensure that all aggregated models are benign (i.e., backdoor-free), then it is obvious that the aggregated global model will also be backdoor-free. 
This intuition can be formally proven if the FL aggregation rule is Byzantine-tolerant. 
To ensure that any backdoor potentially present in the model is eliminated and the aggregated model is benign, a sufficient DP noise level is added to individual local models.
However, since the local models are independent, adding noise to each local model is mathematically equivalent to the case where aggregated noise is added to the global model. This is conceptually equivalent to the conventional centralized setting, for which it has been formally shown that DP noise can eliminate backdoors~\cite{du2020iclr}. In the following, we therefore show that adding DP noise to local models is equivalent to adding `aggregated' DP noise to the global model.}{}

\revisionChanged{
We write the standard deviation of noise for the local models in the form $\sigma_i \gets  \frac{\alpha_i \cdot e_i}{\epsilon} \cdot \sqrt{2ln \frac{1.25}{\delta}}$ where $\alpha_i = \frac{\Delta_i}{e_i}$, $\Delta_i$ and $e_i$ is the sensitivity and the $L_2$ norm of the model $W_i$, respectively.}{}
Mathematically, the FL system with \ourname has: 
\vspace{-0.7em}
\begin{equation}
    \label{eq:local-to-global-noising}
    \begin{aligned}
        G^* &= \frac{1}{n} \Sigma_{i=1}^n W^*_i 
            = \frac{1}{n} [~\Sigma_{i=1}^n ~W_i+N(0,\sigma^2_i)] \\
            &= \frac{1}{n}~\Sigma_{i=1}^2 W_i + \frac{1}{n} \Sigma_{i=1}^n N(0,\sigma_i^2)  \\
            &= \frac{1}{n}~\Sigma_{i=1}^2 W_i + N(0,\frac{1}{n} \Sigma_{i=1}^n \sigma_i^2)  \\
            &= G + N(0,\sigma^2_G) 
    \end{aligned}
\vspace{-0.5em}
\end{equation}
in which $W^*_i$ \revisionChanged{are local models and $G^*$}{} the global model after adding noise $N(0, \sigma^2_i)$. \revisionChanged{Equation~\ref{eq:local-to-global-noising} represents the fact that adding DP noise to each local model (i.e., $W_i+N(0,\sigma^2_i)$) is equivalent to adding an `aggregated' DP noise on the global model (i.e., $G+N(0,\sigma^2_G)$). More specifically, this equivalent Gaussian noise on the global model is the sum of Gaussian noise applied on each local model with a scaling factor $N_G = \frac{1}{n} \Sigma^n_{i=1} N_i$. 
Here, $N_G$ and $N_i$ are random variables with distribution $N(0,\sigma^2_G)$ and $N(0,\sigma^2_i)$, respectively.}{}
As such, we can compute the equivalent noise scale for the global model:
\vspace{-0.6em}
\begin{align}
\label{eq:global-noise}
    \sigma_G^2 &= \frac{1}{n^2} \Sigma^n_{i=1} \sigma^2_i = (\frac{1}{\epsilon}\sqrt{2ln\frac{1.25}{\delta}})^2 \cdot \frac{1}{n^2} \Sigma^n_{i=1}\Delta^2_i \nonumber \\
        &= (\frac{1}{\epsilon}\sqrt{2ln\frac{1.25}{\delta}})^2 \cdot \frac{1}{n^2} \Sigma^n_{i=1}\alpha^2_i e^2_i. 
\vspace{-1.5em}
\end{align}
\revisionChanged{Equation~\ref{eq:global-noise} describes the relation between the DP noise added on \ourname's global model and the DP noise added on each local model. This noise scale relation in Eq.~\ref{eq:global-noise} together with the transformation in Eq. \ref{eq:local-to-global-noising} enable \ourname to provide guaranteed security for the global model against backdoors, thereby addressing Challenge $C_3$ .}{}

\revisionChanged{In Alg. 1, we use the median of Euclidean distances $e_i$ as the upper bound $S_t$ to clip the admitted local models (line 9-11). We hypothesize that the sensitivity of a model $W_i$ is positively correlated with its weight magnitude $|W_i|$ (see Theorem \ref{thm:reduce-noise} for details). In the case of linear models, the sensitivity $\Delta$ has a \textit{linear} relation with the model weight $|\overrightarrow{w}|$ (see Eq. \ref{eq:regression}). Therefore, we use the following approximation: }{}
\vspace{-0.5em}
\begin{equation*}
    \frac{1}{n^2}\Sigma^n_{i=1} \alpha^2_i e^2_i = \frac{1}{n^2}\Sigma^n_{i=1} \Delta^2_i  \approx S^2_t,  
    \vspace{-0.5em}
\end{equation*}
\revisionChanged{where $S_t$ is the weight clipping bound.}{}
\revisionChanged{Having substituted the above approximation into Eq.~\ref{eq:global-noise}, we can compute the noise scale of DP  that \ourname deploys on the global model $N_G$: }{}
\vspace{-0.5em}
\begin{equation}
    \label{eq:sigma_G}
    \sigma_G \approx \frac{S_t}{\epsilon}\sqrt{2ln\frac{1.25}{\delta}}
    \vspace{-0.6em}
\end{equation}

\minorRevision{This concludes the proof of Theorem~\ref{thm:noise-scale}.}{} \qed \\
\revisionChanged{\ourname's adaptive noising step applies the Gaussian noise with the noise scale computed in Eq.~\ref{eq:sigma_G} on the global model for backdoor elimination as shown in Alg. \ref{alg:ouralg}, line 13-14.}{}
Note that \ourname's noising scheme is \textit{adaptive} since the clipping bound $S_t$ is obtained dynamically in each $t^{th}$ epoch. 

\vspace{0.2em}
Next, we present Theorem~\ref{thm:reduce-noise} and justify how \ourname design reduces the derived noise level with \emph{step 3 (S3)} below.

\noindent \textbf{(S3) Clustering and clipping components in \ourname help to reduce the DP noise boundary.}
Recall that \ourname protects the FL system against backdoor attacks using three steps: clustering, clipping, and adding DP noise. 
The overall workflow of \ourname is shown in Fig.~\ref{fig:defense-process}. If multiple backdoors exist in the FL system, the first two steps (clustering and clipping) can remove a subset of backdoors as shown in Fig.~\ref{fig:our-idea}. 
Note that the remaining backdoors are \textit{`closer'} to the benign model updates in terms of both magnitude and direction. This gives us the \textit{intuition} that removing the remaining backdoors by adding DP noise becomes easier (i.e., the noise scale $\sigma_G$ is smaller) after the first two steps of \ourname.  

We can see from Theorem~\ref{thm:noise-scale} that the Gaussian noise scale $\sigma$ required for backdoor resilience increases with the sensitivity of each local model $\Delta_i$.
We describe two characteristics of the model parameter $W$, i.e., direction and magnitude in \sect\ref{sec:idea}. We discuss how these two factors impact the sensitivity of the model defined in \equ~\ref{eq:sens} below.

\vspace{-0.6em}
\begin{theorem}
\label{thm:reduce-noise}
Backdoor models with large angular deviation from benign ones, or with large parameter magnitudes have high sensitivity values $\Delta$.
\end{theorem} 
\vspace{-0.7em}
\revisionAdded{Proving DP-based backdoor security for DNN models is still an open problem, even in the centralized setting. We, therefore, adopt a common approach in literature (e.g., \cite{du2020iclr}) by providing theoretical proof for linear models and validating it for DNNs empirically.}{}

\revisionChanged{\textit{Proof}: for a linear model $f$ where the function output is determined by the inner product of model weight vector $\overrightarrow{w}$ and the data vector $\overrightarrow{x}$, we have}{}
\vspace{-0.5em}
\begin{equation}
\label{eq:regression}
    f(w;~x) = \overrightarrow{w} \cdot \overrightarrow{x} = |w|\cdot|x|\cdot cos\theta, 
    \vspace{-0.5em}
\end{equation}
where $\theta=<\overrightarrow{w},\overrightarrow{x}>$ is the angle between two vectors. 
In this case, it is straightforward to see that if the backdoor attack changes the parameter magnitude $|w|$ or the direction $\theta$ of the model $f$, the resulting poisoned model $f'$ has a large sensitivity value based on the definition in \equ~\ref{eq:sens}.  \qed

This analysis suggests that backdoor models with large angular deviations or with large weight magnitudes have a high sensitivity value $\Delta$. 
Recall that \ourname deploys dynamic clustering (\sect\ref{sec:model-C}) to remove poisoned models with large cosine distances, and employs adaptive clipping (\sect\ref{sec:clipping-noising}) to remove poisoned models with large magnitudes. Therefore, the sensitivity of the remaining backdoor models is lower compared to the one before applying these two steps.
As a result, \emph{\ourname can use a small Gaussian noise to eliminate the remaining backdoors after applying clustering and clipping, which is beneficial for preserving the main task accuracy.} 

We empirically show how the noise scale for backdoor elimination changes after applying each step of \ourname. 
Particularly, we measure the \textit{smallest} Gaussian noise scale $\sigma$ required to defeat \textit{all backdoors} (i.e., $BA=0\%$) in three settings: i) No defense components applied (which is equivalent to the previous DP-based defense~\cite{dwork2014algorithmic,bagdasaryan}); ii) After applying dynamic clustering; 
iii) After applying both dynamic clustering and adaptive clipping (which is the setting of \ourname). 
We conduct this comparison experiment on the \iotTraffic\xspace dataset (cf.~\sect\ref{sec:dataset}). For each communication round, 100 clients are selected where $k=40$ are adversaries. We remove the backdoor by adding Gaussian noise ${N}(0, \sigma^2)$ to the aggregated model. Table~\ref{tab:noise_comp} summarizes the evaluation results in the above three settings. We can observe from the comparison results that the noise scale required to eliminate backdoors decreases after individual deployment of clustering and clipping.
This corroborates the correctness of Theorem~\ref{thm:reduce-noise}. \vspace{-0.4cm}
\begin{table}[t!]
\centering
\caption{Effect of clustering and clipping in \ourname on minimal Gaussian noise level $\sigma$ for backdoor elimination in the NIDS scenario, in terms of Backdoor Accuracy (\ba) and Main Task Accuracy~(\ma).} \label{tab:noise_comp}
\vspace{-0.7em}
\scalebox{0.8}{
\begin{tabular}{|c|rr|rr|rr|} 
\hline
\multirow{2}{*}{\textbf{$\sigma$}} &
\multicolumn{2}{c|}{\begin{tabular}[c]{@{}c@{}}\textbf{Only }\\\textbf{Noising}\end{tabular}} &
\multicolumn{2}{c|}{\begin{tabular}[c]{@{}c@{}}\textbf{After}\\\textbf{Clustering}\end{tabular}} &
\multicolumn{2}{c|}{\begin{tabular}[c]{@{}c@{}}\textbf{After Clustering}\\\textbf{\& Clipping}\end{tabular}} \\
& \multicolumn{1}{c}{BA} & \multicolumn{1}{c|}{MA} & \multicolumn{1}{c}{BA} & \multicolumn{1}{c|}{MA} & \multicolumn{1}{c}{BA} & \multicolumn{1}{c|}{MA}\\
\hline
0.01 & 100.0\%  & 100.0\%  &   0.0\%  & 80.5\%  &   0.0\%  & 100.0\% \\\hline

0.08 &   3.5\%  &  66.7\%  &   0.0\%  &  66.7\%  &   0.0\%  & 100.0\% \\\hline

0.10 &   0.0\%  &  54.2\%  &   0.0\%  &  66.1\%  &   0.0\%  &  87.6\% \\\hline
\end{tabular}
}
\end{table}

\revisionAdded{\subsection{Attack and Data Distribution Assumption}\label{sec:security-assumptions-abscence}}{}
\vspace{-0.15cm}

\revisionAdded{In \ourname, we do not make specific assumptions about the attack and data distribution compared to the existing clustering-based defenses. Let $X= (X_1, \ldots, X_b)$ be a set of distributions of benign models $(W_1, \ldots, W_{n-k})$ where $b \leq n-k$. The deviation in $X$ is caused by the diversity of the data.
Let $X' = (X'_1, \ldots, X'_a)$ be a set of distributions of poisoned models $(W'_1, \ldots, W'_k)$ where $a \leq k$. The deviation in $X'$ is caused by the diversity of the benign data and backdoors (e.g., poisoned data or model crafting).
 Existing works assume that  $X'_i \approx X'_j \quad (\forall i, j: 1 \leq i,j \leq a)$ (see e.g., \cite{fung} or $X' \neq X$  \cite{shen, blanchard}). 
 However, this assumption does not hold in many situations because (i) there can be one or multiple attackers injecting multiple backdoors~\cite{bagdasaryan}, or (ii) the adversary can inject one or several random (honeypot) models having a distribution $X'_r$ that is significantly different from $X \cup (X'\setminus X'_r)$, and (iii) the adversary can control how much the backdoored models deviate from benign ones as discussed in \sect\ref{sec:backdoor_types}. Therefore, approaches that purely divide models into two groups, e.g., K-means~\cite{shen} will incorrectly classify models having distribution $X'_r$ into the malicious group and all remaining models (having distributions drawn from ($X \cup (X' \setminus X'_r)$) into the benign group. As a result, all backdoored models having distributions drawn from ($X'\setminus X'_r$) are classified as benign, as demonstrated in Fig. \ref{fig:cluster-example:kmean}. 
 In contrast, \ourname does not rely on such specific assumptions (the adversary can arbitrarily choose $X'$). 
 If the distribution $X'_i$ of a poisoned model is similar to benign distributions in $X$, \ourname will falsely classify $X'_i$ as being. But if the distribution $X'_j$ of a poisoned model is different from the distributions in $X$, \ourname will identify $X'_j$ as an outlier and classify the associated model as malicious.
 To identify deviating and thus potentially malicious models, \ourname leverages the HDBSCAN algorithm to identify regions of high density in the model space. Any models that are not located in the dense regions will be categorized as outliers, as shown in Fig.~\ref{fig:cluster-example:flguard}. 
 As discussed in \sect\ref{sec:backdoor_types}, \ourname aims to remove models with distributions $X'_j$ that have a higher attack impact compared to models with distribution $X'_i$. It is worth noting, however, that the impact of such remaining backdoored models will be eliminated by the noising component as shown in \sect\ref{sec:noise-proof}}{}
 
\vspace{-0.3em}
\revisionAdded{\noindent\textbf{Striking a balance between accuracy and security:}
Clustering and DP-based approaches affect model accuracy as discussed in \sect\ref{sec:design-challenges} (Challenges $C_2$ and $C_3$). In particular, an approach that aims to maximize the number of filtered malicious models may lead to many false positives, i.e., many benign models being filtered out. Moreover, applying a very low clipping bound or a very high level of injected noise will degrade model accuracy.
To address these problems, \ourname is configured so that the clustering component removes only models with high attack impact rather than all malicious models, i.e., it aims to remove the first backdoor type $W'_1$ as shown in Fig. \ref{fig:idea}. 
In addition, \ourname carefully estimates the clipping bound and noise level to ensure backdoor elimination while preserving model performance. As discussed in \sect\ref{sec:clipping-noising}, the \lnorms of model updates depend on the number of training rounds, dataset types, and type of backdoors. Consequently, the clipping threshold and noise level should be adapted to \lnorms. We therefore apply the median of the \lnorms of model updates as the clipping bound $S_t$ (cf. Lines 9-11 of Alg. \ref{alg:ouralg}). This ensures that $S_t$ is always computed between a benign local model and the global model since we assume that more than 50\% of clients are benign (cf. \sect\ref{sec:adversary-model}). 
Further, estimating noise level based on $S_t$ (cf. Lines 13-14 of Alg. \ref{alg:ouralg}) also provides a noise boundary that ensures that the global model is resilient against backdoors as discussed in \sect\ref{sec:noise-proof}. Moreover, our comparison of potential values for $S_t$ presented in \sect\ref{app:eval-clipping} and \sect\ref{app:eval-noise-adding} shows that the chosen clipping bound and noise level provide the best balance between accuracy and security, i.e., \ourname eliminates backdoor while retaining the global model's performance on the main task. 
}{}

\vspace{-0.5em}
\section{Experimental Setup}
\label{sec:exp-setup}
\vspace{-0.5em}
We conduct all the experiments using the PyTorch deep learning framework~\cite{pytorch} and use the source code provided by Bagdasaryan \etal~\cite{bagdasaryan}, Xie \etal~\cite{xie2020dba} and Wang \etal~\cite{wang2020attack} to implement the attacks. We reimplemented existing defenses to compare them with \ourname.
 
\vspace{-.1em}
\noindent\textbf{Datasets and Learning Configurations.}
\label{sec:dataset}
\minorRevision{\addressedComment{E2}Following recent research on poisoning attacks on FL, we evaluate \ourname in three typical application scenarios: word prediction~\cite{mcmahan2017googleGboard,mcmahan2017aistatsCommunication, mcmahan2018iclrClipping,lin2018iclrCommunication}, image classification~\cite{sheller,sheller2018medical,chilimbi2014osdiDistributed}, and an IoT intrusion detection~\cite{nguyen2019diot, ren2019edgeIoT,samarakoon2018gccV2v,smith2017nipsMultitask} as summarized in Tab. \ref{tab:list-dataset}. Verification of the effectiveness of FLAME against state-of-the-art attacks in comparison to existing defenses (cf. Tab. \ref{tab:effectiveness-attacks} and Tab. \ref{tab:effectiveness-all}) are conducted on these three datasets in the mentioned application scenarios. Experiments for evaluating specific performance aspects of \ourname are performed on the IoT dataset as it represents a very diverse and real-world setting with clear security implications.}{}

\begin{table}[t!]
\vspace{-0.3em}
\centering
  \captionof{table}{Datasets used in our evaluations.}
	\label{tab:list-dataset}
	\vspace{-0.2cm}
	\scalebox{0.92}{
    \EqualTableFontSize{
\begin{tabular}{l|l|l|l|l}
   Application & Datasets & \#Records & Model & \#params \\\hline
    WP & \reddit & 20.6M & LSTM & $\sim$20M\\\hline
    NIDS &\iotTraffic & 65.6M & GRU & $\sim$507k\\\hline
    \multirow{3}{*}{IC} & \cifar & 60k & ResNet-18 Light &$\sim$2.7M\\
    & \mnist& 70k & CNN & $\sim$431k\\
    & \tinyImage & 120k & ResNet-18 & $\sim$11M
	\end{tabular}}}
	
    \vspace{0.1cm}
\end{table}

\noindent\textbf{Evaluation Metrics.}
\minorRevision{We consider a set of metrics for evaluating the effectiveness of backdoor attack and defense techniques as follows: \\
\textbf{BA - Backdoor Accuracy} indicates the accuracy of the model in the backdoor task, i.e., it is the fraction of the trigger set for which the model provides the wrong outputs as chosen by the adversary. 
The adversary aims to maximize \ba, while an effective defense prevents the adversary from increasing it. \\
\textbf{MA - Main Task Accuracy} indicates the accuracy of a model in its main (benign) task. It denotes the fraction of benign inputs for which the system provides correct predictions. 
The adversary aims at minimizing the effect on \ma to reduce the chance of being detected. The defense system should not negatively impact \ma.\\
\textbf{TPR - True Positive Rate} indicates how well the defense identifies poisoned models, i.e., the ratio of the number of models correctly classified as poisoned \revisionAdded{(True Positives - TP) to the total number of models being classified as poisoned: $\tpr = \frac{TP}{TP+FP}$,}{}
\revisionAdded{where FP is False Positives indicating the number of benign clients that are wrongly classified as malicious. 
}{}
\textbf{TNR - True Negative Rate} indicates the ratio of the number of models correctly classified as benign \revisionAdded{(True Negatives - TN) to the total number of benign models: $\tnr = \frac{TN}{TN+FN}$,}{}
\revisionAdded{where FN is False Negatives indicating the number of malicious clients that are wrongly classified as benign. 
}{}}{}

\vspace{-0.5em}
\section{Experimental Results}
\label{sec:exp-result}
\vspace{-1em}
In this section, we evaluate \ourname against backdoor attacks in the literature (\sect\ref{sec:eval-flguard}) \revisionChanged{and demonstrate that our defense mechanism is resilient to adaptive attacks~(\sect\ref{sec:eval-adaptive-attack}). In addition, we show the effectiveness of each of \ournameGen components in \sect\ref{app:eval-each-component} and \ourname overhead in \sect\ref{app:overhead}. Finally, we evaluate the impact of the number of clients (\sect\ref{sec:eval-num-client}) as well as the degree of non-IID data (\sect\ref{sec:eval-iid}).}{}

\vspace{-1.2em}
\subsection{Preventing Backdoor Attacks}
\label{sec:eval-flguard} 
\vspace{-0.3em}
\noindent\textbf{Effectiveness of \ourname}. \revisionChanged{ We evaluate \ourname against the state-of-the-art backdoor attacks called \constrainandscale~\cite{bagdasaryan},  DBA~\cite{xie2020dba}, PGD and \edgeAttack~\cite{wang2020attack}}{} and an untargeted poisoning attack~\cite{fang} (cf.~\sect\ref{app:untargeted}) using the same attack settings as in the original works with multiple datasets. The  results are shown in Tab.~\ref{tab:effectiveness-attacks}. \ourname completely mitigates the \constrainandscale attack ($\ba=0\%$) for all datasets.  Moreover, our defense does not affect the Main Task Accuracy (\ma) of the system as MA reduces by less than $0.4\%$ in all experiments. The DBA attack as well as the \edgeAttack attack~\cite{wang2020attack} are also successfully mitigated ($\ba=3.2\%/4.0\%$). \revisionChanged{Further, \ourname is also effective against PGD attacks ($\ba=0.5~\%$).}{}  It should be noted that suggesting words is a quite challenging task, causing the MA even without attack to be only 22.7\%, aligned with previous work~\cite{bagdasaryan}.
\begin{table}
	\centering
	\caption{Effectiveness of \ourname against \sota attacks for the respective dataset, in terms of Backdoor Accuracy (\ba) and Main Task Accuracy (\ma). All metric values are reported as percentages.
	}
    \vspace{-0.4em}\vspace{-0.15cm}
	\scalebox{0.95}{
	\EqualTableFontSize{
\begin{tabular}{l|l|rr|rr}
                        &\multirow{2}{*}{Dataset}& \multicolumn{2}{c|}{No Defense} & \multicolumn{2}{c}{\ourname} \\
                        \cline{3-6}
                        Attack& & \ba & \ma  &  \ba & \ma \\\hline
                        \multirow{3}{2.7cm}{\mbox{\Constrainandscale}~\cite{bagdasaryan} }&\reddit & 100 & 22.6 & 0 & 22.3\\
                        &\cifar & 81.9& 89.8 & 0&91.9\\
                        &\iotTraffic & 100.0 &100.0 &0 &99.8\\
                        \hline
                        DBA~\cite{xie2020dba} & \cifar &93.8 & 57.4 & 3.2 & 76.2\\
                        \hline
                        \edgeAttack~\cite{wang2020attack} & \cifar & 42.8 & 84.3 & 4.0 & 79.3\\
                        \hline
                        \revisionChanged{PGD~\cite{wang2020attack}}{} & \revisionChanged{\cifar}{} & \revisionChanged{56.1}{} & \revisionChanged{68.8}{} & \revisionChanged{0.5}{} & \revisionChanged{65.1}{}\\
                        \hline
                        Untargeted Poisoning~\cite{fang} & \cifar & - & 46.72& - & 91.31\\
                        

\end{tabular}}}
	\label{tab:effectiveness-attacks}
\end{table}

\minorRevision{We extend our evaluation to various backdoors on three datasets. For NIDS, we evaluate \numberofAttackType different backdoors (Mirai malware attacks) and \numberofDeviceType device types (\numberofDevice IoT devices). The results show that FLAME is able to mitigate all backdoor attacks completely while achieving a high MA=99.8\%. 
We evaluate 5 different word backdoors for WP,  
and 90 different image backdoors for IC, which change the output of a whole class to another class. 
In all cases, \ourname successfully \mbox{mitigates the attack while still preserving the \ma.}}{}

\begin{table}
	\centering
	\caption{Effectiveness of \ourname in comparison to \sota defenses for the \constrainandscale attack on three datasets, in terms of Backdoor Accuracy (\ba) and Main Task Accuracy (\ma). All values are percentages.
	}
    \vspace{-0.6em}
	\label{tab:effectiveness-all}
	\scalebox{0.96}{
	\EqualTableFontSize{
\begin{tabular}{l|rr|rr|rr}
                        \multirow{2}{*}{Defenses}& \multicolumn{2}{c|}{Reddit} & \multicolumn{2}{c|}{\cifar} & \multicolumn{2}{c}{\iotTraffic} \\
                        \cline{2-7}
                        & \ba & \ma  &  \ba & \ma  & \ba & \ma \\\hline
\textit{Benign Setting} &  -    &  22.7 &   -    &   92.2    &  -   &   100.0\\
\textit{No defense}     &  100.0   &  22.6  & 81.9   &   89.8   &   100.0   &   100.0\\
\hline
Krum~\cite{blanchard}                   &   100.0      	&  		   9.6  &  	     100.0  &   	  56.7  &	 100.0	&  84.0\\
FoolsGold~\cite{fung}              & \textbf{0.0}  & 		  \textbf{22.5}  & 	     100.0  &  		  52.3	&		 100.0	&  99.2\\
Auror~\cite{shen}                  &   100.0    	& 		  \textbf{22.5}  &	 100.0	&  		  26.1  &		 100.0	&  96.6\\
AFA~\cite{munoz}                  &   100.0   	&		  22.4  &  \textbf{ 0.0} &   	  91.7  & 		 100.0	&  87.4\\
DP~\cite{dwork2014algorithmic}                      & 		  14.0  &		  18.9	& \textbf{ 0.0} &  		  78.9  &    		  14.8	&  82.3\\

\revisionAdded{Median~\cite{yin2018byzantine}}{}                  & 		  \revisionAdded{\textbf{0.0}}{}  & \revisionAdded{22.0}{}	& \revisionAdded{\textbf{0.0}}{} &  		  \revisionAdded{50.1}{}   &    		  \revisionAdded{0.0}{}	&  \revisionAdded{87.7}{}\\

\cline{1-7}
\ourname                & \textbf{ 0.0} & 22.3  & \textbf{ 0.0} & \textbf{91.9} & \textbf{ 0.0} &  \textbf{99.8}
\end{tabular}}}
\vspace{0.2em}
\end{table}

\label{sect:compexistingdefenses}
\noindent\textbf{Comparison to existing defenses.} \revisionChanged{We compare \ourname to existing defenses: 
Krum~\cite{blanchard}, FoolsGold~\cite{fung}, Auror~\cite{shen}, Adaptive Federated Averaging (AFA) \cite{munoz}, Median~\cite{yin2018byzantine} and a generalized differential privacy (DP) approach~\cite{bagdasaryan, mcmahan2018iclrClipping}.}{} Tab.~\ref{tab:effectiveness-all} shows that \ourname is effective for all 3 datasets, while previous works either fail to mitigate backdoors or reduce the main task accuracy. Krum, FoolsGold, Auror, and AFA do not effectively remove poisoned models and \ba often remains at $100\%$. Also, some defenses make the attack even more successful than without defense. Since they remove many benign updates (cf.~\sect\ref{app:eval-each-component}) but fail to remove a sufficient number of poisoned updates, these defenses increase the \pmr and, therefore, also the impact of the attack. 
\revisionChanged{Some defenses, e.g., Krum~\cite{blanchard}, Auror~\cite{shen} or AFA~\cite{munoz} are not able to handle non-iid data scenarios like Reddit.}{} 
In contrast,  FoolsGold is only effective on the Reddit dataset ($\tpr = 100\%$) because it works well on highly non-independent and identically distributed (non-IID) data (cf.~\sect\ref{sec:related-work}). Similarly, AFA only mitigates backdoors on the \cifar dataset since the data are highly IID (each client is assigned a random set of images) such that the benign models share similar distances to the global model (cf.~\sect\ref{sec:related-work}). Additionally, the model's \ma  is negatively impacted.\\
The DP-based defense is effective, but it significantly reduces \ma. For example, it performs best on the \cifar dataset with $\ba= 0$, but \ma decreases to $78.9\%$ while \ourname increases \ma to $91.9\%$ which is close to the benign setting (no attacks), where  $\ma = 92.2\%$.

\revisionAdded{
\noindent\textbf{Effectiveness of FLAME’s Components.}
Further, we have also conducted an extensive evaluation of the effectiveness of each of FLAME’s components. Due to space limitations, we would like to refer to \sect\ref{app:eval-each-component} for the details.}{}

\vspace{-0.5em}
\subsection{Resilience to Adaptive Attacks}
\label{sec:eval-adaptive-attack}
\vspace{-0.5em}
Given sufficient knowledge about \ourname, an adversary may seek to use adaptive attacks to bypass the defense components. In this section, we analyze such attack scenarios and strategies including \textit{changing the injection strategy}, \textit{model alignment}, and \textit{model obfuscation}.

\noindent\textbf{Changing the Injection Strategy.}
\label{sec:eval-multibackdoor}
The adversary \adversaryClient\xspace may attempt to inject several backdoors simultaneously to execute different attacks on the system in parallel or to circumvent the clustering defense (cf. \sect\ref{sec:existing-attacks}).  
\ourname is also effective against such attacks (cf. Fig.~\ref{fig:cluster-example}). 
To further investigate the resilience of \ourname against such attacks, we conduct two experiments: 1) assigning different backdoors to malicious clients and 2) letting each malicious client inject several backdoors. \minorRevision{\addressedComment{E1} To ensure that each backdoor is injected by a sufficient number of clients, we increased the PMR for this experiment.}{} We conducted these experiments with $n = 100$ clients of which $k = 40$ are malicious on the \iotTraffic\xspace dataset with each type of Mirai attack representing a backdoor. First, we evaluate \ourname for $0, 1, 2$, $4$, and $8$ backdoors, meaning that the number of malicious clients for each backdoor is $0, 40, 20, 10,$ and $5$. Our experimental results show that our approach is effective in mitigating the attacks as $\ba = 0\% \pm 0.0\%$ in all cases, with $\tpr = 95.2\% \pm 0.0\%$, and $\tnr = 100.0\% \pm 0.0\%$.
For the second experiment, 4 backdoors are injected by each of the 40 malicious clients. Also, in this case, the results show that \ourname can completely mitigate the backdoors.

\noindent\textbf{Model Alignment.}
\label{sec:eval-bridging}
Using the same attack parameter values, i.e., \pdr (cf. \sect\ref{sec:existing-attacks}), for all malicious clients can result in high distances between benign and poisoned models. Those high distances can be illustrated as a gap between poisoned and benign models, s.t. the clustering can separate them. Therefore, a sophisticated adversary can generate models that bridge the gap between them such that they are merged to the same cluster in our clustering. We evaluate this attack on the \iotTraffic\xspace dataset for $ k = 80 $ malicious clients and $ n=200 $ clients in total. To remove the gap, each malicious client is assigned a random amount of malicious data, i.e., a random \pdr ranging from $5\%$ to $20\%$. As Tab. \ref{tab:adaptive:briding} shows, \revisionChanged{when we apply model filtering only, our clustering component cannot identify the malicious clients well ($\tpr = 0\%$), resulting in $BA = 100\%$. However, when we apply \ourname, \minorRevision{\addressedComment{E4} although $\tpr$ remains low ($5.68\%$)}, \ourname still mitigates the attack successfully (\ba reduces from $100\%$ to $0\%$). This can be explained by the fact that when the adversary \adversaryClient\xspace tunes malicious updates to be close to the benign ones, the attack's impact is reduced and consequently averaged out by our noising component.}{}

\begin{table}[t!]
\centering
	\caption{Resilience to model alignment attacks in terms of Backdoor Accuracy (\ba), Main Task Accuracy (\ma), True Positive Rate (\tpr), True Negative Rate (\tnr) in percent.}
	\label{tab:adaptive:briding}
	\vspace{-0.6em}
    \EqualTableFontSize{
	\begin{tabular}[!ht]{l|rrrr}
		&  \ba &\ma& \tpr & \tnr\\  	\hline
		Model Filtering & 100.0 & 91.98 & 0.0& 33.04\\
		\ourname  	& \textbf{0.0} & \textbf{100.0} & \textbf{5.68} & \textbf{33.33}\\
\end{tabular}
}
\end{table}

\noindent\textbf{Model Obfuscation.}
\label{sec:eval-obfuscation}
\adversaryClient\xspace can add noise to the poisoned models to make them difficult to detect. However, our evaluation of such an attack on the \iotTraffic\xspace dataset shows that this strategy is not effective.
We evaluate different noise levels to determine a suitable standard deviation for the noise. Thereby, we observe that a noise level of $0.034$ causes the models' cosine distances in clustering to change without significantly impacting \ba. However, \ourname can still efficiently defend this attack: \ba remains at $ 0\% $ and \ma at~100\%.
\vspace{-0.3cm}

\label{sec:eval-increasing-PMR}
\begin{figure}[b]
	\centering{
		\subfloat[IC]{
		\hspace{-0.4cm}
			\includegraphics[width=0.3523\columnwidth]{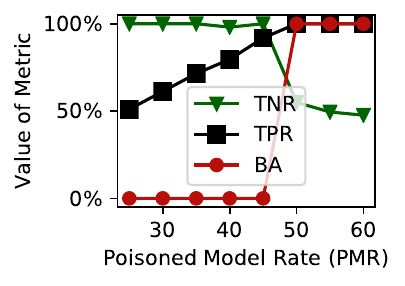}
			\label{fig:eval-PMR:images}
		}\hspace{-0.35cm}
		\subfloat[NIDS]{
			\includegraphics[width=0.34\columnwidth]{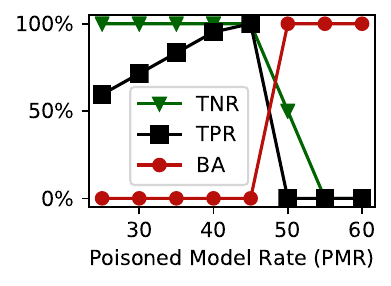}
			\label{fig:eval-PMR:iot}
		}\hspace{-0.35cm}
		\subfloat[WP]{
			\includegraphics[width=0.34\columnwidth]{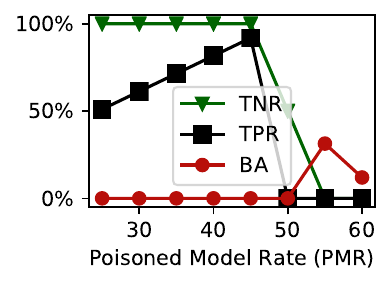}
			\label{fig:eval-PMR:nlp}
		\hspace{-0.4cm}
		}
	}
	\vspace{-1em}
	\caption{Impact of the poisoned model rate $\pmr = \frac{k}{n}$ on the evaluation metrics. $\pmr$ is the fraction of malicious clients $k$ per total clients $n$.}
	\label{fig:eval-PMR}
\end{figure}
\vspace{-0.3em}
\revisionChanged{\subsection{Effect of Number of Clients}}{}
\label{sec:eval-num-client}
\vspace{-0.3em}
\noindent\revisionChanged{\noindent\textbf{Impact of Number of Malicious Clients.}}{}
\revisionChanged{
We assume that the number of benign clients is more than half of all clients (cf.~\sect\ref{sec:existing-attacks}) and our clustering is only expected to be successful when $\pmr=\frac{k}{n}<50\%$ (cf. \sect\ref{sec:model-C}). We evaluate \ourname for different \pmr values. 
Figure~\ref{fig:eval-PMR} shows how \ba, \tpr, and \tnr change in the IC, NIDS, and WP applications for \pmr values from $25\%$ to $60\%$. It shows that \ourname is only effective if $\pmr < 50\%$ so that only benign clients are admitted to the model aggregation ($\tnr = 100\%$) and thus $\ba = 0\%$. However, if $\pmr > 50\%$, \ourname fails to mitigate the attack because the majority of poisoned models will be included resulting in low $\tnr$. Interestingly, \ourname accepted all models for $\pmr=50\%$ ($\tpr = 0\%$ and $\tnr = 100\%$).  
For the IC application, 
since the IC data are non-IID, poisoned models are not similar. Therefore, some poisoned models were excluded from the cluster resulting in a high \tpr even for \pmrs higher than 50\%. However, the majority of poisoned models were selected resulting in the drop in the \tnr.}{} 

\begin{figure}[t]
	\centering{
		\subfloat[Image Classification]{
			\includegraphics[clip,trim={0 0.25cm 0 0.2cm},width=0.9\columnwidth]{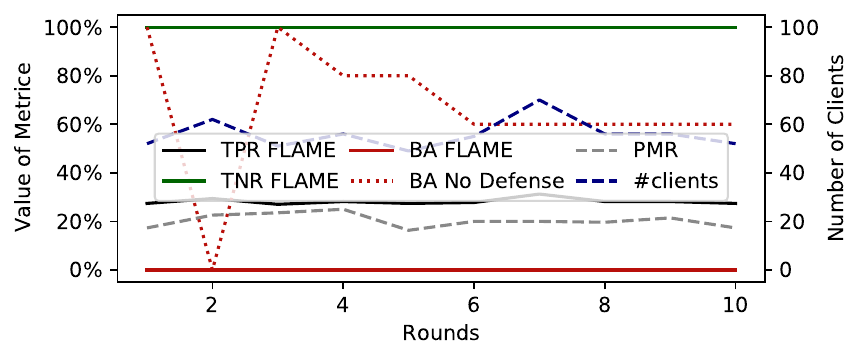}
			\label{fig:eval-randomClientNumber:images}
		}\vspace{-0.35cm}\\
		\subfloat[Network Intrusion Detection]{
	        \includegraphics[clip,trim={0 0.25cm 0 0.2cm},width=0.9\columnwidth]{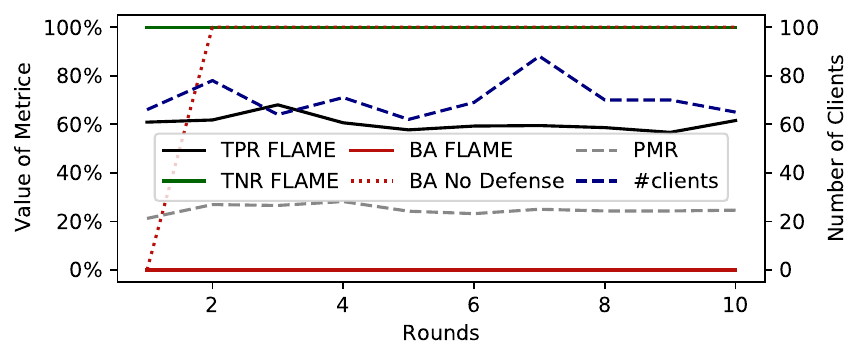}
	    	\label{fig:eval-randomClientNumber:iot}
		}
	}
	\vspace{-1em}
	\caption{Impact of the number of clients on \ourname}
	\label{fig:eval-randomClientNumber}
\end{figure}

\textbf{Varying number of clients in different training rounds.} In general, \ourname is a round-independent defense, i.e., it does not use information from previous rounds such as which clients were excluded in which rounds. Therefore, \ourname will not be affected if the number of clients or number of malicious clients varies as long as the majority of clients remain benign. To demonstrate this, we simulate realistic scenarios in which clients can join and drop out dynamically. We conducted an experiment where during each round, the total number of available clients is randomly selected. As the result, the number of malicious clients will also be random. \minorRevision{\addressedComment{E1} In this experiment, we used a population of 100 clients in total, out of which 25 are malicious. In each round, a random number (from 60 to 90) of clients are selected, so that the fraction of malicious clients ($\pmr$) varies in each round.
Figure~\ref{fig:eval-randomClientNumber} shows the experimental results. One can see that the proportion of malicious clients ($\pmr$) does not affect the effectiveness of \ourname, i.e., the backdoor is completely removed ($BA = 0\%$) in every round. Since all poisoned models are detected, their negative effect on the aggregated model is removed. Therefore, the MA with \ourname is better than the one without defense, and is almost always 100~\% aligned with the results in Tab. \ref{tab:effectiveness-all}.}{}

\vspace{-0.5em}
\subsection{Impact of the Degree of non-IID Data}
\label{sec:eval-iid}
\vspace{-0.5em}

Since clustering is based on measuring differences between benign and malicious updates, the distribution of data among clients might affect our defense. \revisionAdded{We conduct two experiments for both \Constrainandscale and \edgePGDAttack on the \cifar dataset. For Reddit and IoT datasets, changing the degree of non-IID data is not meaningful since the data have a natural distribution as every client obtains data from different Reddit users or traffic chunks from different IoT devices. Following previous works~\cite{fang,wang2020attack}, we vary the degree of non-IID data \degNonIID by changing the fraction of images belonging to a specific class assigned to clients. In particular, we divide the clients into 10 groups corresponding to the 10 classes of \cifar. The clients of each group are assigned a fixed fraction of \degNonIID of the images from its designated image class, while the rest of the images will be assigned to it at random. Consequently, the data distribution is random, i.e., completely IID if $\degNonIID = 0\%$ (all images are randomly assigned) and completely non-IID if $\degNonIID = 100\%$ (a client only gets images from its designated class).}\\
\begin{figure}[tb]
\captionsetup[subfloat]{captionskip=0pt}
	\centering{
	
		\subfloat[\Constrainandscale]{
			\includegraphics[width=0.75\columnwidth]{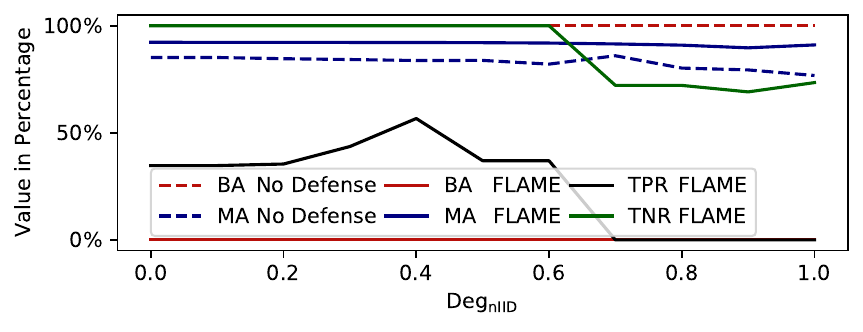}
			\label{fig:eval-clustering-iid:images}
		}
	\vspace{-0.3cm}
		\subfloat[\edgePGDAttack]{
			\includegraphics[width=0.75\columnwidth]{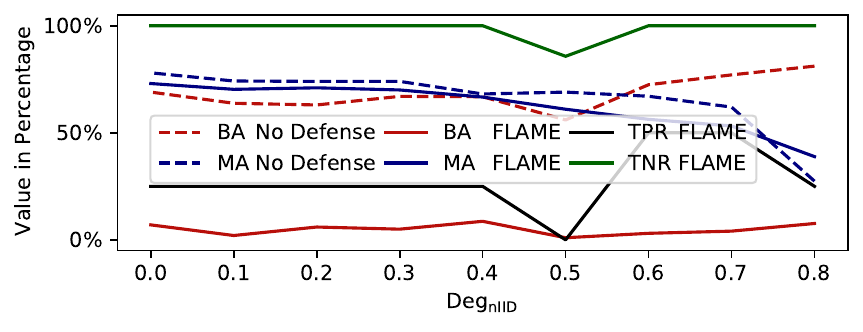}
			\label{fig:eval-clustering-iid:edge}
		}
	}
	\setlength{\belowcaptionskip}{-6pt}
	\vspace{-1em}
	\caption{Impact of degree of non-IID data on \ourname for \constrainandscale using the \degNonIID and for the \edgePGDAttack attack using the $\alpha$ parameter of the Dirichlet distribution.}
	\label{fig:eval-clustering-iid}
	 \vspace{0.6em}
\end{figure}

\revisionAdded{Figure~\ref{fig:eval-clustering-iid:images} shows the evaluation results for the \constrainandscale attacks. Although \ourname does not detect the poisoned models for very non-IID scenarios, it still mitigates the attack as the BA remains $0\%$ for all values of \degNonIID. For low \degNonIID, \ourname effectively identifies the poisoned models ($TNR=100\%$) and the MA remains on almost the same level as without defense.}{} 
\revisionAdded{As shown in Fig.~\ref{fig:eval-clustering-iid:edge}, \ourname also mitigates the \edgePGDAttack attack effectively for all $\alpha$ values of the Dirichlet distribution and the MA also stays on the same level as without defense.}{}
However, since not all poisoned models are detected, a higher $\sigma$ is determined dynamically to mitigate the \constrainandscale backdoor, resulting in a slightly reduced MA for $\degNonIID \geq 0.7$ (MA is $91.9\%$ for $\degNonIID=0.6$, and is reduced to $91.0\%$ for $\degNonIID=1.0$).
Note that Fig.~\ref{fig:eval-clustering-iid} shows the evaluation results in a training round $t$ where the global model $G_t$ is close to convergence \cite{bagdasaryan}, thus even though the TNR decreases with a large value of \degNonIID, the drop of MA with \ourname is not substantial.

\vspace{-0.5em}
\section{Privacy-preserving Federated Learning}
\label{sec:privacy}

\vspace{-1em}

A number of attacks on FL have been proposed that aim to infer from parameters of a model the presence of a specific training sample in the training dataset (\emph{membership inference attacks})~\cite{Pyrgelis,shokri,melis2019exploiting}, properties of training samples (\emph{property inference attacks})~\cite{Ganju,melis2019exploiting}, try to assess the proportion of samples of a specific class in the data (\emph{distribution estimation attacks})~\cite{wang2019arxivEavesdrop}. Inference attacks by the aggregator~\adversaryServer\xspace are significantly stronger, as \adversaryServer\xspace has access to the local models~\cite{nasr2019comprehensive} and can also link gained information to a specific user, while the global model anonymizes the individual contributions. Therefore, enhanced privacy protection for FL is needed that prohibits access to the local model updates. 

\mypar{Adversary Model (privacy)}
In this adversary type, $\adversaryServer$ attempts to infer sensitive information about clients' data~$D_i$ from their model updates~$W_i$
~\cite{Pyrgelis,shokri, Ganju, melis2019exploiting} by maximizing the information  $\phi_i = \Call{Infer}{W_i}$ that $\adversaryServer$ gains about the data $D_i$ of client~$i$ by inferring from its corresponding model~$W_i$.


\mypar{Deficiencies of existing defenses}
Generally, there are two approaches to protect the privacy of clients’ data: differential privacy (DP;~\cite{dwork2014algorithmic}) and cryptographic techniques such as homomorphic encryption~\cite{gentry09} or multi-party computation~\cite{Demmler2015}. DP is a statistical approach that 
can be efficiently implemented, but it can only offer high privacy protection at the cost of a significant loss in accuracy due to the noise added to the models~\cite{BatchCrypt,aono2017privacy}.
In contrast, cryptographic techniques provide strong privacy guarantees as well as high accuracy at the cost of reduced efficiency. 

\begin{figure}[t]
	\centering
	\includegraphics[width=1.0\columnwidth]{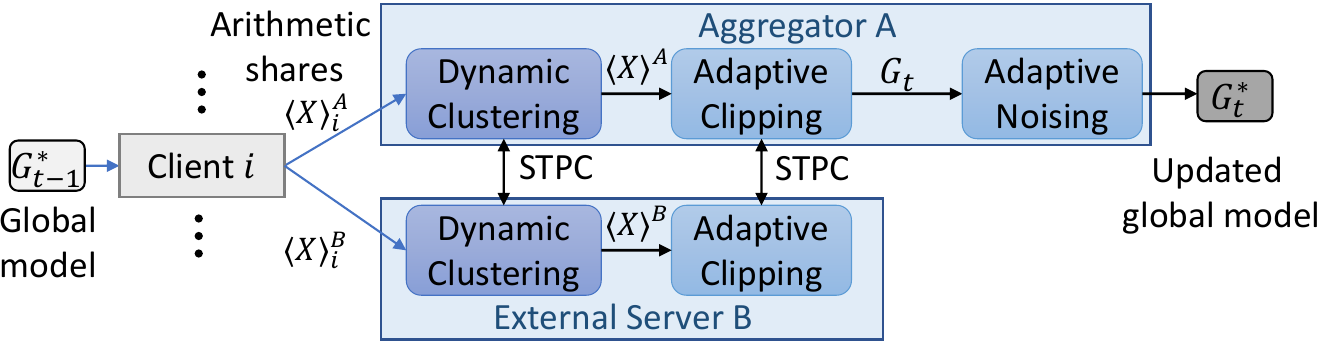}
	\vspace{-2em}
	\caption{Overview of private \ourname in round $t$ using Secure-Two-Party Computation (STPC). \vspace{0.2cm}}
  \label{fig:mpc-fl} 
\end{figure}

\noindent \textbf{Private \ourname.}
\label{subsec:privateFL}
To securely implement \ourname using STPC, we use an optimized combination of three prominent STPC techniques as implemented with state-of-the-art optimizations in the ABY framework~\cite{Demmler2015}. 
\HCHANGED{
Fig.~\ref{fig:mpc-fl} shows an overview of private~\ourname. It involves $n$ clients and two non-colluding servers, called aggregator $A$ and external server~$B$. Each client~$i\in \{1,...,n\}$ splits the parameters of its local update $W_i$ into two Arithmetic shares $\langle X\rangle^A_i$ and $\langle X\rangle^B_i$, such that $W_i = \langle X\rangle _i^A +\langle X \rangle _i^B $, and sends $\langle X\rangle _i^A$ to $A$ and $\langle X \rangle _i^B $ to $B$. $A$ and $B$ then privately compute the new global model via STPC. 
We co-design the distance calculation, clustering, adaptive clipping, and aggregation of \ourname (cf.~Alg.~\ref{alg:ouralg}) of \ourname as efficient STPC protocols.
To further improve performance, we approximate \hdbscan with the simpler DBSCAN \cite{boz2021} to avoid the construction of the minimal spanning tree in \hdbscan as it is very expensive to realize with STPC. 
\minorRevision{See~\sect\ref{sec:evalstpc} for more details on private \ourname evaluation of its accuracy and performance.}{}}

\vspace{0.5em}

\vspace{-0.5em}

\vspace{-0.5em}
\section{Related Work}
\label{sec:related-work}

\vspace{-1em}

In general, existing backdoor defenses can roughly be divided into two main categories. The first one aims to distinguish malicious updates and benign updates by 1) clustering model updates \cite{shen, fung, blanchard, diakonikolas2019sever, li2019abnormal, khazbakmlguard, li2020learning}, 2) changing aggregation rules \cite{yin2018byzantine, guerraoui2018hidden}, and 3) using root dataset \cite{andreina2020baffle}. The second category is based on differential privacy techniques \cite{sun2019can, bagdasaryan}. 
Next, we will discuss these points in detail.

\noindent\textbf{Clustering model updates.} Several backdoor defenses, such as Krum~\cite{blanchard}, AFA~\cite{munoz}, and Auror~\cite{shen}, aim at separating benign and malicious model updates. However, they only work under specific assumptions about the underlying data distributions, e.g., Auror and Krum assume that data of benign clients are iid. In contrast, FoolsGold and AFA \cite{munoz} assume that benign data are non-iid. In addition, FoolsGold assumes that manipulated data are iid. As a result, these defenses are only effective under specific circumstances (cf.~\sect\ref{sec:eval-flguard}) and cannot handle the simultaneous injection of multiple backdoors (cf.~\sect\ref{sec:model-C}). 
Moreover, such defenses cannot detect stealthy attacks, e.g., where the adversary constrains their poisoned updates within benign update distribution such as \Constrainandscale attacks \cite{bagdasaryan}. In contrast, \ourname does not make any assumption about the data distribution, clipping, and noising components can also mitigate stealthy attacks, and \ourname can defend against \mbox{injection of multiple backdoors~(cf.~\sect\ref{sec:model-C}).}

\noindent\textbf{Changing aggregation rules.}
Instead of using FedAvg \cite{mcmahan2017aistatsCommunication}, Yin \etal~\cite{yin2018byzantine} and Guerraoui \etal~\cite{guerraoui2018hidden} propose using the median parameters from all local models as the global model parameters, i.e., $G_t = \Call{Median}{W^t_1, \ldots, W^t_n}$. However, the adversary can bypass it by injecting stealthy models like $W'_3$ (cf. Fig. \ref{fig:backdoor_types}), in which the parameters of the poisoned model will be selected to be incorporated into the global model. \revisionChanged{Further, our evaluation in \sect\ref{sec:eval-flguard} shows that Median also reduces the performance of the model significantly.}{}

\noindent\minorRevision{\addressedComment{B2 and D1}\textbf{Using root data.} Although FLTrust~\cite{cao2020fltrust} can defend against byzantine clients (with arbitrary behavior) and detect poisoning attacks, including backdoors, it requires the aggregator to have access to a benign root dataset.}{}
Baffle~\cite{andreina2020baffle} utilizes clients using their own data to evaluate the performance of the aggregated model to detect backdoors. 
However, this approach has several limitations. Firstly, the backdoor triggers are only known to the attacker. One cannot ensure that the benign clients would have such trigger data to activate the backdoor. 
Secondly, Baffle does not work in a non-IID data scenario with a small number of clients as clients cannot distinguish deficits in model performance due to the backdoor from lack of data. 

\noindent\textbf{Differential Privacy-based approaches.} 
Clipping and noising are known techniques to achieve differential privacy (DP)~\cite{dwork2014algorithmic}. However, directly applying these techniques to defend against backdoor attacks is not effective because they significantly decrease the Main Task Accuracy (\sect\ref{sec:eval-flguard}) \cite{bagdasaryan}. \ourname tackles this by i) identifying and filtering out potential poisoned models that have a high attack impact (cf.~\sect\ref{sec:model-C}), and ii) eliminating the residual poison with an appropriate adaptive clipping bound and noise level, such that the Main Task Accuracy is retained (cf.~\sect\ref{sec:clipping-noising}).

\section{\verfiveChanged{Revisiting Attacks Claiming to Bypass FLAME}}
\label{sec:new-attacks}
\verfiveChanged{Since we presented our \ourname paper at the USENIX Security Conference in 2022 \cite{nguyen2022Usenix}, it has sparked considerable interest among researchers, garnering over 100 citations according to Google Scholar \cite{flame-gscholar}. When creating this document, two attacks have been highlighted that allege to circumvent \ourname \cite{li2023sp, Xu2022ACSAC}. We promptly responded by conducting an exhaustive investigation into these attacks, uncovering notable errors in their approach. They made misleading claims and seemed to misinterpret \ourname's core design principles. To resolve this issue, we initiated a dialogue with the respective authors, leading to an acknowledgment and ensuing correction of assertions by Xu et al. \cite{Xu2022ACSAC}. In the subsequent sections, we explore these studies in greater detail.}

\verfiveChanged{\noindent\textbf{Xu et al.} \cite{Xu2022ACSAC} propose a distributed backdoor attack (DBA) targeting federated Graph Neural Networks (GNN) training. The authors assert their proposed attacks can circumvent \ourname \cite{nguyen2022Usenix}. However, upon discussing this with the authors of \cite{Xu2022ACSAC}, it becomes clear that they have misinterpreted our approach and failed to incorporate all key components of \ourname in their understanding. Specifically, they have only accounted for the clustering aspect of \ourname while neglecting the Adaptive Clipping and Noising components, both of which are essential for backdoor mitigation. Acknowledging this, the authors have revised their claim in the recent version of their paper published on arXiv \cite{Xu2022arXiv}.}

\verfiveChanged{ \noindent\textbf{Li et al. \cite{li2023sp}} introduce \thrdfed, a multifaceted backdoor attack composed of three elements: Constrain Loss, Noise Masks, and Decoy Model. Among these, the Constrain Loss component is crafted to target \ourname. However, it solely aims to counteract the Clipping defense facet of \ourname, as outlined in \sect{VI} of \cite{li2023sp}. This implies that the authors did not account for the Clustering and Noising elements intrinsic to our defense design. At its core, \thrdfed seeks to limit the loss during the backdoor model training phase, striving to keep the backdoor objective closely aligned with the global model in terms of \lnorm (Euclidean distance). This alignment is achieved by manipulating the hyper-parameter $\beta$, meaning \thrdfed's goal is to ensure the \lnorm of the backdoor model falls under the \lnorm-based clipping threshold.}
\verfiveChanged{In \cite{li2023sp}, however, the authors neither discuss nor assess how hyper-parameter $\beta$ could be selected to ensure \lnorm{s} of backdoored models fall below our dynamic Clipping boundary. It is important to note that because the Clipping threshold in \ourname is dynamically recalculated with each federated training iteration, it's highly unlikely that a particular $\beta$ value can ensure the \lnorm{s} of tampered models stay below our dynamic Clipping threshold while still maintaining the attack's effectiveness. This is particularly true once Gaussian noise is factored in.  As we highlighted in Section \ref{sec:idea},  we are aware of such tampered models, working on the assumption that an adversary can carefully manipulate the training process (see Constrain-and-scale attack \cite{bagdasaryan}). 
Consequently, we argue that \thrdfed cannot counteract \ourname. To empirically support this argument, we have carried out a comprehensive replication study of \thrdfed using the exact source code provided by the \thrdfed authors \cite{3dfedCode}, strictly adhering to the experimental setup detailed in the 3DFed paper \cite{li2023sp}, as described in Section \ref{sec:app-3dfed}. As anticipated, contrary to the results presented in the \thrdfed paper, specifically Figures 10c, 10h, and 13c \cite{li2023sp}, our results demonstrate \ourname's effective mitigation of 3DFed attacks across all datasets used in \thrdfed experiments (see Table \ref{tab:rep-3dfed}).}

\verfiveChanged{In conclusion, Table \ref{tab:new-attacks} underscores the significant discrepancy between how recent attacks interpret \ourname and the actual design principles of \ourname. Importantly, none of these approaches consider the Adaptive Noising defense, which plays a crucial role in eliminating remaining backdoor updates. This highlights the essentiality of comprehending design principles when devising an attack against \ourname.}

\begin{table}[ht!]
\centering 
\caption{\verfiveChanged{Analysis of Recent Attack Strategies Against \ourname.}}
\label{tab:new-attacks}	
    
\begin{tabular}{|c|cc|}
\hline
\multirow{2}{*}{\begin{tabular}[c]{@{}c@{}}\ourname defense\\ mechanisms\end{tabular}} & \multicolumn{2}{c|}{Consideration in attack design}          \\ \cline{2-3} 
                                                                                    & \multicolumn{1}{c|}{Xu et al.\cite{Xu2022ACSAC}} & Li et al. \cite{li2023sp} \\ \hline
Dynamic Clustering                                                                  & \multicolumn{1}{c|}{Yes}             & No               \\ \hline
Adaptive Clipping                                                                   & \multicolumn{1}{c|}{No}              & Yes              \\ \hline
Adaptive Noising                                                                    & \multicolumn{1}{c|}{No}              & No               \\ \hline
\end{tabular}

\end{table}

\vspace{-0.5em}
\section{Conclusion}
\label{sec:conclusion}
\vspace{-1em}
In this paper, we introduce \ourname, a resilient aggregation framework for FL that eliminates the impact of backdoor attacks while maintaining the performance of the aggregated model on the main task. We propose a method to approximate the amount of noise that needs to be injected into the global model to neutralize backdoors. Furthermore, in combination with our dynamic clustering and adaptive clipping, \ourname can significantly reduce the noise scale for backdoor removal and thus preserve the benign performance of the global model. In addition, we design, implement, and benchmark efficient secure two-party computation protocols for \ourname to ensure the privacy of clients' training data and to impede inference attacks on client updates.

\section*{Acknowledgments}

This research was funded by the Deutsche Forschungsgemeinschaft (DFG) SFB-1119 CROSSING/236615297, the European Research Council (ERC, grant No. 850990 PSOTI), the EU H2020 project SPATIAL (grant No. 101021808), GRK 2050 Privacy \& Trust/251805230, HMWK within ATHENE project, NSF-TrustHub (grant No. 1649423), SRC-Auto (2019-AU-2899), Huawei OpenS3 Lab, and Intel Private AI Collaborative Research Center. We thank the anonymous reviewers and the shepherd, Neil Gong, for constructive reviews and comments.


{\small
\bibliographystyle{plain}
\bibliography{reference}

\begin{thebibliography}{10}

\bibitem{flame-gscholar}
Flame: Taming backdoors in federated learning.
\newblock
  \url{https://scholar.google.de/citations?view_op=view_citation&hl=en&user=Ro2m0z0AAAAJ&citation_for_view=Ro2m0z0AAAAJ:GnPB-g6toBAC}.

\bibitem{redditDataset}
Reddit dataset, 2017.
\newblock
  \url{https://bigquery.cloud.google.com/dataset/fh-bigquery:reddit_comments}.

\bibitem{pytorch}
Pytorch, 2019.
\newblock \url{https://pytorch.org}.

\bibitem{3dfedCode}
3dfed source code.
\newblock \url{https://github.com/haoyangliASTAPLE/3DFed}, 2023.

\bibitem{abadi2016ccsDifferential}
Martin Abadi, Andy Chu, Ian Goodfellow, H.~Brendan McMahan, Ilya Mironov, Kunal
  Talwar, and Li~Zhang.
\newblock Deep learning with differential privacy.
\newblock In {\em CCS}. {ACM}, 2016.

\bibitem{andreina2020baffle}
Sebastien Andreina, Giorgia~Azzurra Marson, Helen Möllering, and Ghassan
  Karame.
\newblock {BaFFLe: Backdoor Detection via Feedback-based Federated Learning}.
\newblock In {\em ICDCS}, 2021.

\bibitem{antonakakis2017usenixMirai}
Manos Antonakakis, Tim April, Michael Bailey, Matt Bernhard, Elie Bursztein,
  Jaime Cochran, Zakir Durumeric, J.~Alex Halderman, Luca Invernizzi, Michalis
  Kallitsis, Deepak Kumar, Chaz Lever, Zane Ma, Joshua Mason, Damian Menscher,
  Chad Seaman, Nick Sullivan, Kurt Thomas, and Yi~Zhou.
\newblock {Understanding the Mirai Botnet}.
\newblock In {\em {USENIX} Security}, 2017.

\bibitem{aono2017privacy}
Yoshinori Aono, Takuya Hayashi, Lihua Wang, and Shiho Moriai.
\newblock {Privacy-preserving Deep Learning via Additively Homomorphic
  Encryption}.
\newblock In {\em TIFS}, 2017.

\bibitem{bagdasaryan}
Eugene Bagdasaryan, Andreas Veit, Yiqing Hua, Deborah Estrin, and Vitaly
  Shmatikov.
\newblock {How To Backdoor Federated Learning}.
\newblock In {\em AISTATS}, 2020.

\bibitem{baruch}
Moran Baruch, Gilad Baruch, and Yoav Goldberg.
\newblock {A Little Is Enough: Circumventing Defenses For Distributed
  Learning}.
\newblock In {\em NIPS}, 2019.

\bibitem{blanchard}
Peva Blanchard, El~Mahdi El~Mhamdi, Rachid Guerraoui, and Julien Stainer.
\newblock {Machine Learning with Adversaries: Byzantine Tolerant Gradient
  Descent}.
\newblock In {\em NIPS}, 2017.

\bibitem{boz2021}
Beyza Bozdemir, Sébastien Canard, Orhan Ermis, Helen Möllering, Melek Önen,
  and Thomas Schneider.
\newblock Privacy-preserving density-based clustering.
\newblock In {\em ASIACCS}, 2021.

\bibitem{campello2013HDBSCAN}
Ricardo J. G.~B. Campello, Davoud Moulavi, and Joerg Sander.
\newblock {Density-Based Clustering Based on Hierarchical Density Estimates}.
\newblock In {\em {Pacific-Asia Conference on Knowledge Discovery and Data
  Mining}}, 2013.

\bibitem{cao2020fltrust}
Xiaoyu Cao, Minghong Fang, Jia Liu, and Neil~Zhenqiang Gong.
\newblock Fltrust: Byzantine-robust federated learning via trust bootstrapping.
\newblock In {\em NDSS}, 2021.

\bibitem{chilimbi2014osdiDistributed}
Trishul Chilimbi, Yutaka Suzue, Johnson Apacible, and Karthik Kalyanaraman.
\newblock {Project Adam: Building an Efficient and Scalable Deep Learning
  Training System}.
\newblock In {\em {USENIX} Operating Systems Design and Implementation}, 2014.

\bibitem{Demmler2015}
Daniel Demmler, Thomas Schneider, and Michael Zohner.
\newblock {ABY - A Framework for Efficient Mixed-Protocol Secure Two-Party
  Computation}.
\newblock In {\em NDSS}, 2015.

\bibitem{diakonikolas2019sever}
Ilias Diakonikolas, Gautam Kamath, Daniel Kane, Jerry Li, Jacob Steinhardt, and
  Alistair Stewart.
\newblock Sever: A robust meta-algorithm for stochastic optimization.
\newblock In {\em ICML}, 2019.

\bibitem{Doshi2018DDoSDetection}
Rohan Doshi, Noah Apthorpe, and Nick Feamster.
\newblock {Machine Learning DDoS Detection for Consumer Internet of Things
  Devices}.
\newblock In {\em arXiv preprint:1804.04159}, 2018.

\bibitem{du2020iclr}
Min Du, Ruoxi Jia, and Dawn Song.
\newblock Robust anomaly detection and backdoor attack detection via
  differential privacy.
\newblock In {\em ICLR}, 2020.

\bibitem{dwork2014algorithmic}
Cynthia Dwork and Aaron Roth.
\newblock {The Algorithmic Foundations of Differential Privacy}.
\newblock In {\em Foundations and Trends in Theoretical Computer Science},
  2014.

\bibitem{ester1996density}
Martin Ester, Hans-Peter Kriegel, J{\"o}rg Sander, Xiaowei Xu, et~al.
\newblock {A Density-Based Algorithm for Discovering Clusters in Large Spatial
  Databases with Noise.}
\newblock In {\em KDD}, 1996.

\bibitem{fang}
Minghong {Fang}, Xiaoyu {Cao}, Jinyuan {Jia}, and Neil {Zhenqiang Gong}.
\newblock {Local Model Poisoning Attacks to Byzantine-Robust Federated
  Learning}.
\newblock In {\em {USENIX} Security}, 2020.

\bibitem{fereidooni2022fedcri}
Hossein Fereidooni, Alexandra Dmitrienko, Phillip Rieger, Markus Miettinen,
  Ahmad-Reza Sadeghi, and Felix Madlener.
\newblock {FedCRI: Federated Mobile Cyber-Risk Intelligence}.
\newblock In {\em NDSS}, 2022.

\bibitem{fereidooni2021safelearn}
Hossein Fereidooni, Samuel Marchal, Markus Miettinen, Azalia Mirhoseini, Helen
  M{\"o}llering, Thien~Duc Nguyen, Phillip Rieger, Ahmad-Reza Sadeghi, Thomas
  Schneider, Hossein Yalame, et~al.
\newblock Safelearn: secure aggregation for private federated learning.
\newblock In {\em 2021 IEEE Security and Privacy Workshops (SPW)}, pages
  56--62. IEEE, 2021.

\bibitem{fung}
Clement Fung, Chris~JM Yoon, and Ivan Beschastnikh.
\newblock The limitations of federated learning in sybil settings.
\newblock In {\em {{RAID}}}, 2020.
\newblock originally published as arxiv:1808.04866.

\bibitem{Ganju}
Karan Ganju, Qi~Wang, Wei Yang, Carl~A Gunter, and Nikita Borisov.
\newblock {Property Inference Attacks on Fully Connected Neural Networks Using
  Permutation Invariant Representations}.
\newblock In {\em CCS}, 2018.

\bibitem{gentry09}
Craig Gentry.
\newblock {\em A Fully Homomorphic Encryption Scheme}.
\newblock PhD thesis, Stanford University, Stanford, CA, USA, 2009.

\bibitem{guerraoui2018hidden}
Rachid Guerraoui, S{\'e}bastien Rouault, et~al.
\newblock The hidden vulnerability of distributed learning in byzantium.
\newblock In {\em ICML}. {PMLR}, 2018.

\bibitem{he2016deep}
Kaiming He, Xiangyu Zhang, Shaoqing Ren, and Jian Sun.
\newblock {Deep Residual Learning for Image Recognition}.
\newblock In {\em CVPR}, 2016.

\bibitem{herwig2019ndssHajime}
Stephen Herwig, Katura Harvey, George Hughey, Richard Roberts, and Dave Levin.
\newblock {Measurement and Analysis of Hajime, a Peer-to-Peer IoT Botnet}.
\newblock In {\em NDSS}, 2019.

\bibitem{huang2019arxivMedical}
Li~Huang, Yifeng Yin, Zeng Fu, Shifa Zhang, Hao Deng, and Dianbo Liu.
\newblock {LoAdaBoost: Loss-Based AdaBoost Federated Machine Learning on
  medical Data}.
\newblock In {\em arXiv preprint:1811.12629}, 2018.

\bibitem{khazbakmlguard}
Youssef Khazbak, Tianxiang Tan, and Guohong Cao.
\newblock Mlguard: Mitigating poisoning attacks in privacy preserving
  distributed collaborative learning.
\newblock In {\em International Conference on Computer Communications and
  Networks (ICCCN)}. IEEE, 2020.

\bibitem{kolias2017Miraivarients}
Constantinos Kolias, Georgios Kambourakis, Angelos Stavrou, and Jeffrey Voas.
\newblock {DDoS in the IoT: Mirai and Other Botnets}.
\newblock In {\em IEEE Computer}, 2017.

\bibitem{krizhevsky2009learning}
Alex Krizhevsky and Geoffrey Hinton.
\newblock {Learning Multiple Layers of Features from Tiny Images}.
\newblock Technical report, 2009.

\bibitem{lecun1998gradient}
Y.~Lecun, L.~Bottou, Y.~Bengio, and P.~Haffner.
\newblock Gradient-based learning applied to document recognition.
\newblock {\em Proceedings of the IEEE}, 86(11), 1998.

\bibitem{li2023sp}
Haoyang Li, Qingqing Ye, Haibo Hu, Jin Li, Leixia Wang, Chengfang Fang, and Jie
  Shi.
\newblock 3dfed: Adaptive and extensible framework for covert backdoor attack
  in federated learning.
\newblock In {\em 2023 IEEE Symposium on Security and Privacy (SP)}, pages
  1893--1907, 2023.

\bibitem{li2019abnormal}
Suyi Li, Yong Cheng, Yang Liu, Wei Wang, and Tianjian Chen.
\newblock Abnormal client behavior detection in federated learning.
\newblock {\em arXiv preprint arXiv:1910.09933}, 2019.

\bibitem{li2020learning}
Suyi Li, Yong Cheng, Wei Wang, Yang Liu, and Tianjian Chen.
\newblock Learning to detect malicious clients for robust federated learning.
\newblock {\em arXiv preprint arXiv:2002.00211}, 2020.

\bibitem{lin2018iclrCommunication}
Yujun Lin, Song Han, Huizi Mao, Yu~Wang, and William~J. Dally.
\newblock {Deep Gradient Compression: Reducing the Communication Bandwidth for
  Distributed Training}.
\newblock In {\em ICLR}, 2018.

\bibitem{mcinnes2017accelerated}
Leland McInnes and John Healy.
\newblock Accelerated hierarchical density based clustering.
\newblock In {\em Data Mining Workshops (ICDMW), 2017 IEEE International
  Conference on}. IEEE, 2017.

\bibitem{mcinnes2017hdbscan}
Leland McInnes, John Healy, and Steve Astels.
\newblock hdbscan: Hierarchical density based clustering.
\newblock {\em The Journal of Open Source Software}, 2(11):205, 2017.

\bibitem{mcmahan2017aistatsCommunication}
Brendan McMahan, Eider Moore, Daniel Ramage, Seth Hampson, and
  Blaise~Ag{\"{u}}era y~Arcas.
\newblock {Communication-Efficient Learning of Deep Networks from Decentralized
  Data}.
\newblock In {\em AISTATS}, 2017.

\bibitem{mcmahan2017googleGboard}
Brendan McMahan and Daniel Ramage.
\newblock {Federated learning: Collaborative Machine Learning without
  Centralized Training Data}.
\newblock Google AI, 2017.

\bibitem{mcmahan2018iclrClipping}
H.~Brendan McMahan, Daniel Ramage, Kunal Talwar, and Li~Zhang.
\newblock {Learning Differentially Private Language Models Without Losing
  Accuracy}.
\newblock In {\em ICLR}, 2018.

\bibitem{melis2019exploiting}
Luca Melis, Congzheng Song, Emiliano De~Cristofaro, and Vitaly Shmatikov.
\newblock {Exploiting Unintended Feature Leakage in Collaborative Learning}.
\newblock In {\em IEEE S\&P}, 2019.

\bibitem{munoz}
Luis {Mu{\~n}oz-Gonz{\'a}lez}, Kenneth~T. {Co}, and Emil~C. {Lupu}.
\newblock {Byzantine-Robust Federated Machine Learning through Adaptive Model
  Averaging}.
\newblock In {\em arXiv preprint:1909.05125}, 2019.

\bibitem{nasr2019comprehensive}
M.~{Nasr}, R.~{Shokri}, and A.~{Houmansadr}.
\newblock Comprehensive privacy analysis of deep learning: Passive and active
  white-box inference attacks against centralized and federated learning.
\newblock In {\em IEEE S\&P}, 2019.

\bibitem{nguyen2019diot}
Thien~Duc Nguyen, Samuel Marchal, Markus Miettinen, Hossein Fereidooni,
  N.~Asokan, and Ahmad{-}Reza Sadeghi.
\newblock {{D\"{I}oT}: A Federated Self-learning Anomaly Detection System for
  IoT}.
\newblock In {\em ICDCS}, 2019.

\bibitem{nguyen2022Usenix}
Thien~Duc Nguyen, Phillip Rieger, Huili Chen, Hossein Yalame, Helen
  M{\"o}llering, Hossein Fereidooni, Samuel Marchal, Markus Miettinen, Azalia
  Mirhoseini, Shaza Zeitouni, Farinaz Koushanfar, Ahmad-Reza Sadeghi, and
  Thomas Schneider.
\newblock {FLAME}: Taming backdoors in federated learning.
\newblock In {\em 31st USENIX Security Symposium (USENIX Security 22)}, pages
  1415--1432, Boston, MA, August 2022. USENIX Association.

\bibitem{nguyen2020diss}
Thien~Duc Nguyen, Phillip Rieger, Markus Miettinen, and Ahmad-Reza Sadeghi.
\newblock {Poisoning Attacks on Federated Learning-Based IoT Intrusion
  Detection System}.
\newblock In {\em Workshop on Decentralized IoT Systems and Security}, 2020.

\bibitem{Pyrgelis}
Apostolos Pyrgelis, Carmela Troncoso, and Emiliano De~Cristofaro.
\newblock {Knock Knock, Who’s There? Membership Inference on Aggregate
  Location Data}.
\newblock In {\em NDSS}, 2018.

\bibitem{ren2019edgeIoT}
Jianji Ren, Haichao Wang, Tingting Hou, Shuai Zheng, and Chaosheng Tang.
\newblock {Federated Learning-Based Computation Offloading Optimization in Edge
  Computing-Supported Internet of Things}.
\newblock In {\em IEEE Access}, 2019.

\bibitem{samarakoon2018gccV2v}
Sumudu {Samarakoon}, Mehdi {Bennis}, Walid {Saad}, and Merouane {Debbah}.
\newblock {Federated Learning for Ultra-Reliable Low-Latency V2V
  Communications}.
\newblock In {\em GLOBCOM}, 2018.

\bibitem{sheller}
Micah {Sheller}, Anthony {Reina}, Brandon {Edwards}, Jason {Martin}, and
  Spyridon {Bakas}.
\newblock {Federated Learning for Medical Imaging}.
\newblock In {\em Intel AI}, 2018.

\bibitem{sheller2018medical}
Micah {Sheller}, Anthony {Reina}, Brandon {Edwards}, Jason {Martin}, and
  Spyridon {Bakas}.
\newblock {Multi-Institutional Deep Learning Modeling Without Sharing Patient
  Data: A Feasibility Study on Brain Tumor Segmentation}.
\newblock In {\em Brain Lesion Workshop}, 2018.

\bibitem{shen}
Shiqi Shen, Shruti Tople, and Prateek Saxena.
\newblock {Auror: Defending Against Poisoning Attacks in Collaborative Deep
  Learning Systems}.
\newblock In {\em ACSAC}, 2016.

\bibitem{shokri}
Reza Shokri, Marco Stronati, Congzheng Song, and Vitaly Shmatikov.
\newblock {Membership Inference Attacks Against Machine Learning Models}.
\newblock In {\em IEEE S\&P}, 2017.

\bibitem{sivanathan2018UNSWdata}
Arunan Sivanathan, Hassan~Habibi Gharakheili, Franco Loi, Adam Radford, Chamith
  Wijenayake, Arun Vishwanath, and Vijay Sivaraman.
\newblock {Classifying IoT Devices in Smart Environments Using Network Traffic
  Characteristics}.
\newblock In {\em TMC}, 2018.

\bibitem{smith2017nipsMultitask}
Virginia Smith, Chao-Kai Chiang, Maziar Sanjabi, and Ameet~S Talwalkar.
\newblock {Federated Multi-Task Learning}.
\newblock In {\em NIPS}, 2017.

\bibitem{soltan2018usenixBlackIoT}
Saleh Soltan, Prateek Mittal, and Vincent Poor.
\newblock {BlackIoT: IoT Botnet of High Wattage Devices Can Disrupt the Power
  Grid}.
\newblock In {\em {USENIX} Security}, 2018.

\bibitem{sun2019can}
Ziteng Sun, Peter Kairouz, Ananda~Theertha Suresh, and H~Brendan McMahan.
\newblock Can you really backdoor federated learning?
\newblock {\em arXiv preprint arXiv:1911.07963}, 2019.

\bibitem{wang2020attack}
Hongyi Wang, Kartik Sreenivasan, Shashank Rajput, Harit Vishwakarma, Saurabh
  Agarwal, Jy-yong Sohn, Kangwook Lee, and Dimitris Papailiopoulos.
\newblock Attack of the tails: Yes, you really can backdoor federated learning.
\newblock In {\em NeurIPS}, 2020.

\bibitem{wang2019arxivEavesdrop}
Lixu {Wang}, Shichao {Xu}, Xiao {Wang}, and Qi~{Zhu}.
\newblock {Eavesdrop the Composition Proportion of Training Labels in Federated
  Learning}.
\newblock In {\em arXiv preprint:1910.06044}, 2019.

\bibitem{xie2020dba}
Chulin Xie, Keli Huang, Pin-Yu Chen, and Bo~Li.
\newblock {DBA: Distributed Backdoor Attacks against Federated Learning}.
\newblock In {\em ICLR}, 2020.

\bibitem{Xu2022ACSAC}
Jing Xu, Rui Wang, Stefanos Koffas, Kaitai Liang, and Stjepan Picek.
\newblock More is better (mostly): On the backdoor attacks in federated graph
  neural networks.
\newblock In {\em Annual Computer Security Applications Conference (ACSAC)
  2022}, 2022.

\bibitem{Xu2022arXiv}
Jing Xu, Rui Wang, Stefanos Koffas, Kaitai Liang, and Stjepan Picek.
\newblock More is better (mostly): On the backdoor attacks in federated graph
  neural networks, 2022.

\bibitem{yin2018byzantine}
Dong Yin, Yudong Chen, Ramchandran Kannan, and Peter Bartlett.
\newblock Byzantine-robust distributed learning: Towards optimal statistical
  rates.
\newblock In {\em ICML}. {PMLR}, 2018.

\bibitem{BatchCrypt}
Chengliang Zhang, Suyi Li, Junzhe Xia, Wei Wang, Feng Yan, and Yang Liu.
\newblock {BatchCrypt: Efficient Homomorphic Encryption for Cross-Silo
  Federated Learning}.
\newblock In {\em USENIX ATC}, 2020.

\end{thebibliography}
}

\appendix

\vspace{-1.2em}
\section{Datasets and Learning Configurations}
\label{app:datasets}
\vspace{-0.6em}

\noindent\textbf{Word Prediction (WP).} 
\minorRevision{
We use the Reddit dataset of November 2017 \cite{redditDataset} with the same settings as state-of-the-art works \cite{bagdasaryan,mcmahan2017aistatsCommunication, mcmahan2018iclrClipping} for comparability. In particular, each user in the dataset with at least 150 posts and not more than 500 posts is considered as a client. This results in \numprint{80000} clients' datasets with sizes between 298 and \numprint{32660} words. 

The model consists of two LSTM layers and a linear output layer \cite{bagdasaryan, mcmahan2017aistatsCommunication}. 
To be comparable to the attack setting in~\cite{bagdasaryan}, we evaluate \ourname on five different backdoors, each with a different trigger sentence corresponding to a chosen output. 

\noindent\textbf{Image Classification (IC).} For image classification, we use mainly the \cifar dataset~\cite{krizhevsky2009learning}, a standard benchmark dataset for image classification, in particular for FL~\cite{mcmahan2017aistatsCommunication} and backdoor attacks~\cite{bagdasaryan, baruch, munoz}. 
It consists of \numprint{60000} images of 10 different classes. The adversary aims at changing the predicted label of one class of images to another class of images. 
We use a lightweight version of the ResNet18 model~\cite{he2016deep} with 4 convolutional layers with max-pooling and batch normalization~\cite{bagdasaryan}. The experimental setup consists of 100 clients and uses a PMR of $20\%$. 
In addition to the \cifar dataset, we also evaluate \ourname's effectiveness on two further datasets for image classification. 
The \textit{\mnist} dataset consists of \numprint{70000} handwritten digits~\cite{lecun1998gradient}. The learning task is to classify images to identify digits. The adversary poisons the model by mislabeling labels of digit images before using it for training \cite{shen}. We use a convolutional neural network (CNN) with 431000 parameters. The \textit{\tinyImage} \footnote{https://tiny-imagenet.herokuapp.com} consists of 200 classes and each class has 500 training images, 50 validation images, and 50 test images. We used ResNet18~\cite{he2016deep} model.


\noindent\textbf{Network Intrusion Detection System (NIDS).} \label{sec:iot-dataset}
We test backdoor attacks on IoT anomaly-based intrusion detection systems that often represent critical security applications~\cite{antonakakis2017usenixMirai, herwig2019ndssHajime,Doshi2018DDoSDetection,soltan2018usenixBlackIoT,kolias2017Miraivarients, nguyen2019diot, nguyen2020diss}. Here, the adversary aims at causing incorrect classification of anomalous traffic patterns, e.g., generated by IoT malware, as benign patterns. Based on the FL anomaly detection system \mbox{\diot~\cite{nguyen2019diot}}, we use three datasets called DIoT-Benign, DIoT-Attack, and UNSW-Benign~\cite{nguyen2019diot, sivanathan2018UNSWdata} from real-world home and office deployments (four homes and two offices located in Germany and Australia). 
DIoT-Attack contains the traffic of 5 anomalously behaving IoT devices, infected by the Mirai malware~\cite{nguyen2019diot}.  
Moreover, we collected a fourth IoT dataset containing communication data from \numberofDeviceType typical IoT devices (including IP cameras and power plugs) in three different smart home settings and an office setting. Following~\cite{nguyen2019diot}, we extracted device-type-specific datasets capturing the devices' communication behavior. 
We simulate the FL setup by splitting each device type's dataset among several clients (from 20 to 200). Each client has a training dataset corresponding to three hours of traffic measurements containing samples of roughly \numprint{2000}-\numprint{3000} communication packets.
The learning model \mbox{consists of 2 GRU layers and a fully connected layer.}
}{} 

\vspace{-0.6em}

\section{Effectiveness of \ourname's Components}   
\label{app:eval-each-component}
\vspace{-0.4em}

\begin{table}[tb]
	\centering
	\caption{Effectiveness of the clustering component, in terms of True Positive Rate (\tpr) and True Negative Rate (\tnr), of \ourname in comparison to existing defenses for the \constrainandscale attack on three datasets. All values are in percentage and the best results of the defenses are marked in bold.}
	\vspace{-0.8em}
	\label{tab:effectiveness-clustering}
    \scalebox{1.}{
    \EqualTableFontSize{
\begin{tabular}{l|rr|rr|rr}
                        \multirow{2}{*}{Defenses}& \multicolumn{2}{c|}{Reddit} & \multicolumn{2}{c|}{\cifar} & \multicolumn{2}{c}{\iotTraffic} \\
                        &  TPR & TNR & TPR & TNR & TPR & TNR \\\hline
Krum                   &  9.1  & 0.0    & 	    		8.2  & 			 0.0      &		   24.2  & 			 0.0\\
FoolsGold              &\textbf{100.0}  &\textbf{100.0} & 	    			0.0  &			90.0&		   32.7  &			84.4\\
Auror                  &   0.0  &90.0&					0.0  &			90.0&		    0.0  &			70.2\\
AFA                  &    0.0  &88.9& \textbf{100.0} & \textbf{100.0}&	    4.5  & 			69.2\\
\cline{1-7}
\ourname                &  22.2 & \textbf{100.0}& 23.8			 & 		86.2& 	   \textbf{59.5}  &		   \textbf{100.0}
\end{tabular}
}}
\vspace{0.3em}
\end{table}
\vspace{-0.6em}
\subsection{Effectiveness of the Clustering Component}
\label{app:eval-clustering} 
\vspace{-0.5em}

We show the results for the clustering component in Tab.~\ref{tab:effectiveness-clustering}. As shown there, our filtering achieves $\tnr = 100\%$ for the \reddit and \iotTraffic\xspace datasets, \ie, \ourname only selects benign models in this attack setting. 
Recall that the goal of clustering is to filter out the poisoned models with high attack impact, i.e., not necessarily all poisoned models (cf. \sect\ref{sec:high-level}). This allows \ourname to defend backdoor attacks effectively, even if not all poisoned models are filtered. For example, although for the \cifar dataset in Tab.~\ref{tab:effectiveness-clustering} the TNR is not 100~\% ($86.2\%$), the attack is still mitigated by the noising component, such that the BA is 0~\% (cf. Tab.~\ref{tab:effectiveness-all}).
\begin{figure}[t]
\vspace{-0.5cm}
	\centering{
		\subfloat[BA for static bound]{
			\includegraphics[width=0.425\columnwidth]{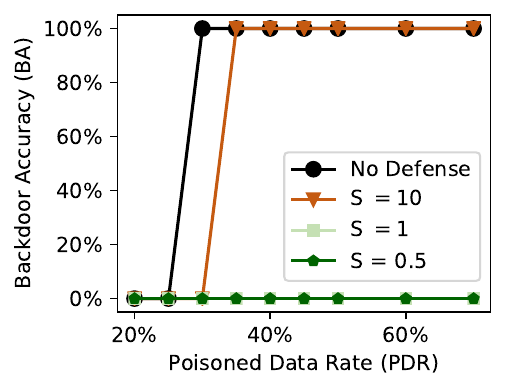}
			\label{fig:static-bound-BA}
		}
		\subfloat[MA for static bound]{
			\includegraphics[width=0.425\columnwidth]{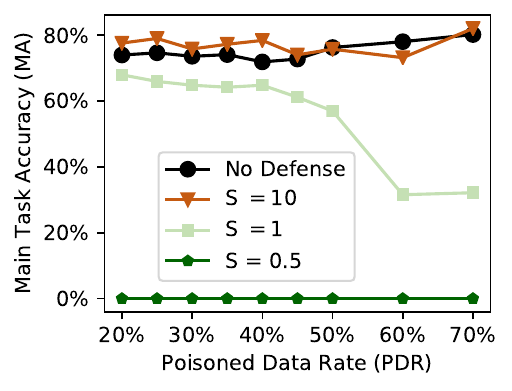}
			\label{fig:static-bound-MA}
		}\\\vspace{-0.25cm}
		\subfloat[BA for dynamic bound]{
			\includegraphics[width=0.425\columnwidth]{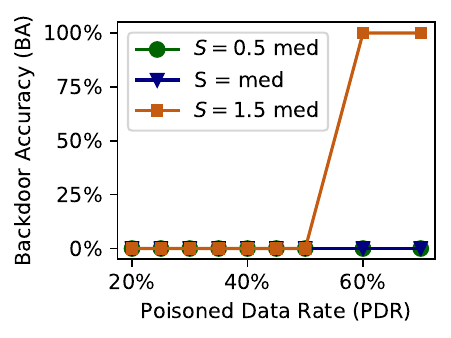}
			\label{fig:dynamic-bound-BA}
		}
		\subfloat[MA for dynamic bound]{
			\includegraphics[width=0.425\columnwidth]{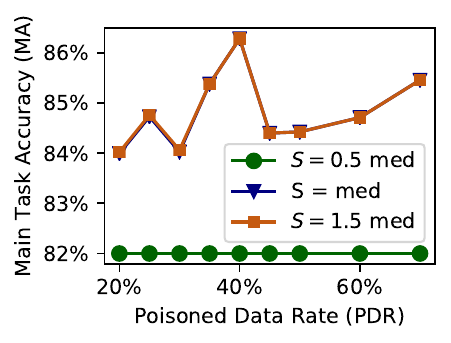}
			\label{fig:dynamic-bound-MA}
		}
	}
	\vspace{-0.6em}
	\caption{Effectiveness of \ourname's clipping bound in terms of Backdoor Accuracy (\ba) and Main Task Accuracy (\ma). 
	$ S $ is the clipping bound and $med$ the \lnorm median.}
	\label{fig:dynamic-clipping-bound}
	\vspace{-0.1em}
\end{figure}

\vspace{-0.8em}
\subsection{Effectiveness of Clipping} 
\label{app:eval-clipping}
\vspace{-0.5em}
\minorRevision{Fig.~\ref{fig:dynamic-clipping-bound} demonstrates the effectiveness of \ourname's dynamic clipping where S is the median of \lnorms 
compared to a static clipping bound~\cite{bagdasaryan} and different choices for a dynamic clipping boundary (i.e., median, half of median, median multiplied by $1.5$). The experiments are conducted for the \iotTraffic\xspace dataset, which is \noniid.}{} Fig.~\ref{fig:static-bound-BA} and Fig.~\ref{fig:static-bound-MA} show that a small static bound $S = 0.5$ is effective to mitigate the attack ($\ba = 0\%$), but \ma drops to $0\%$ rendering the model useless. Moreover, a higher static bound like $S = 10$ is ineffective as $\ba = 100\%$ if the Poisoned Data Rate (\pdr) $\geq 35\%$. In contrast, \ourname's dynamic clipping threshold performs significantly better as \ba consistently remains at $0\%$ while \ma remains high (cf. Fig.~\ref{fig:dynamic-bound-BA} and Fig.~\ref{fig:dynamic-bound-MA}). 

\vspace{-1em}
\subsection{Effectiveness of Adding Noise}
\label{app:eval-noise-adding}
\begin{figure}[t]
\vspace{-0.45cm}
	\centering{
		\subfloat[Image Classification]{
			\includegraphics[width=0.49\columnwidth]{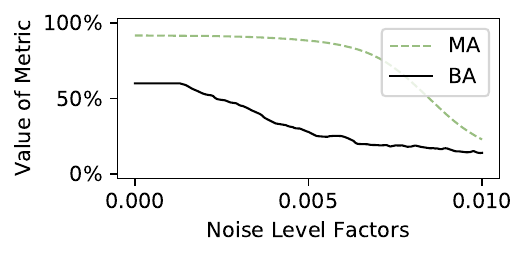}
			\label{fig:eval-component-parameter-noiselevels:images}
		}
		\subfloat[Network Intrusion Detection]{
			\includegraphics[width=0.49\columnwidth]{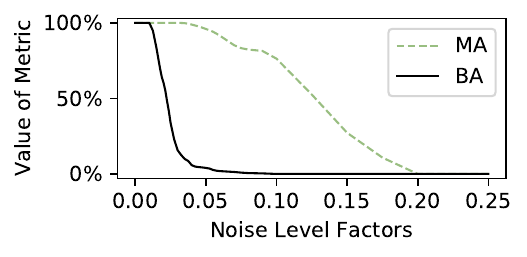}
			\label{fig:eval-component-parameter-noiselevels:iot}
		}
	}	
	\caption{Impact of different noise level factors on the Backdoor Accuracy (\ba) and Main Task Accuracy (\ma).}
	\label{fig:eval-component-parameter-noiselevels}
\end{figure}

\vspace{-0.6em}
Fig.~\ref{fig:eval-component-parameter-noiselevels} shows the impact of adding noise to the intermediate global models with respect to different noise level factors $\lambda$ to determine the standard deviation of the noise $\sigma$ dynamically based on the median \lnorm of the updates $S_t$ as $\sigma =\lambda S_t$. As it can be seen, increasing $\lambda$ reduces the \ba, but it also negatively impacts the performance of the model in the main task (\ma). Therefore, the noise level must be dynamically tuned and combined with the other defense components to optimize the overall success of the defense. 
The noise level factor is determined by $\lambda=\frac{1}{\epsilon} \sqrt{2ln\frac{1.25}{\delta}}$ for $(\epsilon,\delta)$-DP. We use standard DP parameters and set $\epsilon=3705$ for IC, $\epsilon=395$ for the NIDS and $\epsilon=4191$ for the NLP scenario. Accordingly, $\lambda = 0.001$ for IC and NLP, and $\lambda = 0.01$ for the NIDS scenario. \minorRevision{\addressedComment{B1} The DP budget is dependent on the considered dataset scenario. It is determined based on the median of the dataset sizes of the clients and the size of the model used. It can thus be empirically determined by the aggregator. Analogous to determining the clipping boundary S (cf. \ref{sec:clipping-noising}), using the median ensures that the used dataset size is within the range of benign values.}{}



\vspace{-0.8em}
\section{Na\"ive Combination}
\label{app:naive-combination}
\vspace{-0.6em}
Furthermore, we test a na\"ive combination of the defense components by stacking clipping and adding noise (using a fixed clipping bound of 1.0 and a standard deviation of 0.01 as in \cite{bagdasaryan}) on top of a clustering component using K-means. However, this na\"ive approach still allows a \ba of 51.9\% and a \ma of 60.24\%, compared to a \ba of 0.0\% and a \ma of 89.87\% of \ourname in the same scenario for the \cifar dataset. Based on our evaluations in \sect\ref{sect:compexistingdefenses}, it becomes apparent that \ourname's dynamic nature goes beyond previously proposed defenses that consist of static baseline ideas, which \ourname significantly optimizes, extends, and automates to offer a comprehensive dynamic and private defense against sophisticated backdoor attacks.

\vspace{-1.3em}
\section{Overhead of \ourname}
\label{app:overhead}
\vspace{-0.9em}
\minorRevision{ We evaluated \ourname for 6 different device types from the IoT dataset. 
In this experiment, only benign clients participated and the model was randomly initialized. The highest observed overhead was 4 additional rounds. In average, $1.67$ additional training rounds were needed to achieve at least 99\% of the \ma that was achieved without applying the defense, i.e., \ourname does not prevent the model from converging.}{}

\vspace{-0.6em}
\section{HDBSCAN}
\label{sec:HDBSCAN}
\vspace{-0.6em}

\minorRevision{
HDBSCAN~\cite{campello2013HDBSCAN} is a density-based clustering technique that classifies data samples in different clusters without predefined the maximum distance and the number of clusters. In the following, we describe HDBSCAN in detail, following the implementation of McInnes \etal~~\cite{mcinnes2017accelerated, mcinnes2017hdbscan}. However, we focus on the behavior of HDBSCAN for the parameters that \ourname uses, i.e., when \texttt{min\_cluster\_size=$\nicefrac{N}{2}+1$} and \texttt{min\_samples=1}, e.g., because of the choice for \texttt{min\_cluster\_size} we skip parts that deal with multiple clusters.}{} 
\minorRevision{HDBSCAN first uses the given distances to build a minimal spanning tree (MST), where the vertices represent the individual data points and the edges are weighted by the distances between the respective points. 
Then it uses the MST to build a binary tree where the leaf nodes represent the vertices of the MST and the non-leaf nodes represent the edges of the MST. For this, first, all vertices are considered as separate trees (of size 1). For this, first, all vertices are considered as separate trees (of size 1) and then, starting from the edge with the lowest weight, iteratively the trees are merged by creating a non-leaf-node for each edge of the MST and set the (previously not connected) subtrees containing the endpoints of the edge as children for the new node (represented by calling the function \texttt{make\_binary\_tree}.}{} 
\minorRevision{In the next step, HDBSCAN collects all nodes of the binary tree as candidates, that cover at least $\nicefrac{N}{2}+1$ data points. 
Since only non-leaf nodes fulfill the requirement of covering at least $\nicefrac{N}{2}+1$ data points, each cluster candidate is based on a node, representing an edge in the MST. It uses the weight of the edge and the number of covered points to calculate a so-called stability value.}{} 
\minorRevision{Then HDBSCAN uses the stability value to determine the cluster candidate with the most homogeneous density and returns this candidate as majority cluster. 
Finally, it assigns the cluster label to all data points inside this cluster and labels all points outside of this cluster as noise. 
}{}

\vspace{-0.8em}
\section{Effectiveness of \ourname against untargeted poisoning attacks}
\label{app:untargeted}
\vspace{-0.6em}
\minorRevision{Another attack type related to backdooring is \emph{untargeted poisoning}~\cite{fang, blanchard, baruch}. Unlike backdoor attacks that aim to incorporate specific backdoor functionalities, untargeted poisoning aims at rendering the model unusable. 
The adversary uses crafted local models with low Main Task Accuracy to damage the global model $G$. Fang at el.\cite{fang} propose such an attack bypassing \sota defenses. 
Although we do not focus on untargeted poisoning, our approach intuitively defends it since, in principle, this attack also trade-offs attack impact against stealthiness. To evaluate the effectiveness of \ourname against this attack, 
we test the Krum-based attack proposed by \cite{fang} on \ourname. 
Since \cite{fang}'s evaluation uses image datasets, we evaluate \ourname's resilience against it with CIFAR-10. 
The evaluation results show that although the attack significantly damages the model by reducing \ma from $92.16\%$ to $46.72\%$, \ourname can successfully defend against it and \ma remains at $91.31\%$.  
}{}

\vspace{-0.8em}
\section{Performance of Private \ourname}
\label{sec:evalstpc}
\vspace{-0.6em}

\HCHANGED{For our implementation, we use the STPC framework ABY~\cite{Demmler2015} which implements the three sharing types, including \sota optimizations and flexible conversions and the open-source privacy-preserving DBSCAN by Bozdemir et al.~\cite{boz2021}. All STPC results are averaged over 10 experiments and run on two separate servers with Intel Core i9-7960X CPUs with 2.8 GHz and 128 GB RAM connected over a 10 Gbit/s LAN with 0.2 ms RTT.}

\minorRevision{
\noindent\textbf{Approximating \hdbscan by \dbscan.}
We measure the effect of approximating \hdbscan by \dbscan including the binary search for the neighborhood parameter $\epsilon$. The results show that our approximation has a negligible loss of accuracy. For some applications, the approximation 
even performs slightly better than the standard \ourname, e.g., for \cifar, private \ourname correctly filters all poisoned models, while standard \ourname accepts a small number ($\tnr = 86.2\%$), which is still sufficient to achieve $\ba = 0.0\%$. 


\noindent\textbf{Runtime of Private \ourname.}
 We evaluate the runtime in seconds per training iteration of the cosine distance, Euclidean distance + clipping + model aggregation, and clustering steps of Alg.~\ref{alg:ouralg} in standard (without STPC) and in private \ourname (with STPC). The results show that private \ourname causes a significant overhead on the runtime by a factor of up to three orders of magnitude compared to the standard (non-private) \ourname. However, even if we consider the largest model (Reddit) with $K=100$ clients, we have a total server-side runtime of 22\,081.65 seconds ($\approx$ 6 hours) for a training iteration with STPC. Such runtime overhead would be acceptable to maintain privacy, especially since mobile phones, which would be a typical type of clients in FL~\cite{mcmahan2017aistatsCommunication}, are not always available and connected so that there will be delays in synchronizing clients' model updates in FL. These delays can then also be used to run STPC. Furthermore, achieving provable privacy by using STPC may even motivate more clients to contribute to FL in the first place and provide more data. 
}{}

\begin{table}[t]
\centering 
\caption{\verfiveChanged{Efficacy of FLAME in mitigating 3DFed across various datasets and pre-trained models. *) MA is not reported while FLAME is deployed. Models were pre-trained following the specifications outlined in the configuration file \cite{3dfedCode}.}}
\label{tab:rep-3dfed}	

\begin{tabular}{|c|ccc|}
\hline
Dataset                                                                   & \multicolumn{1}{c|}{\begin{tabular}[c]{@{}c@{}}Pre-trained\\ Model\\ @Epoch\end{tabular}} & \multicolumn{1}{c|}{\begin{tabular}[c]{@{}c@{}}Averaged MA\\ for 50 exp. \\ (in \%)\end{tabular}} & \begin{tabular}[c]{@{}c@{}}Averaged BA\\ for 50 exp. \\ (in \%)\end{tabular} \\ \hline
\multirow{7}{*}{CIFAR-10}                                                 & \multicolumn{3}{c|}{3DFed paper}                                                                                                                                                                                                                                             \\ \cline{2-4} 
                                                                          & \multicolumn{1}{c|}{200}                                                                  & \multicolumn{1}{c|}{\begin{tabular}[c]{@{}c@{}}Not\\ reported*\end{tabular}}                      & 99.89                                                                        \\ \cline{2-4} 
                                                                          & \multicolumn{3}{c|}{Our replication study}                                                                                                                                                                                                                                   \\ \cline{2-4} 
                                                                          & \multicolumn{1}{c|}{200}                                                                  & \multicolumn{1}{c|}{\textit{80.8}}                                                                & \textbf{53.9}                                                                \\
                                                                          & \multicolumn{1}{c|}{400}                                                                  & \multicolumn{1}{c|}{\textit{86.9}}                                                                & \textbf{9.9}                                                                 \\
                                                                          & \multicolumn{1}{c|}{600}                                                                  & \multicolumn{1}{c|}{\textit{87.0}}                                                                & \textbf{9.9}                                                                 \\
                                                                          & \multicolumn{1}{c|}{800}                                                                  & \multicolumn{1}{c|}{\textit{86.9}}                                                                & \textbf{9.9}                                                                 \\ \hline
\multirow{7}{*}{MNIST}                                                    & \multicolumn{3}{c|}{3DFed paper}                                                                                                                                                                                                                                             \\ \cline{2-4} 
                                                                          & \multicolumn{1}{c|}{10**}                                                                 & \multicolumn{1}{c|}{\begin{tabular}[c]{@{}c@{}}Not\\ reported*\end{tabular}}                      & 96                                                                           \\ \cline{2-4} 
                                                                          & \multicolumn{3}{c|}{Our replication study}                                                                                                                                                                                                                                   \\ \cline{2-4} 
                                                                          & \multicolumn{1}{c|}{10}                                                                   & \multicolumn{1}{c|}{\textit{98.9}}                                                                & \textbf{19.1}                                                                \\
                                                                          & \multicolumn{1}{c|}{20}                                                                   & \multicolumn{1}{c|}{\textit{99.1}}                                                                & \textbf{20.4}                                                                \\
                                                                          & \multicolumn{1}{c|}{30}                                                                   & \multicolumn{1}{c|}{\textit{99.1}}                                                                & \textbf{24.0}                                                                \\
                                                                          & \multicolumn{1}{c|}{50}                                                                   & \multicolumn{1}{c|}{\textit{99.1}}                                                                & \textbf{29.2}                                                                \\ \hline
\multirow{7}{*}{\begin{tabular}[c]{@{}c@{}}Tiny-\\ ImageNet\end{tabular}} & \multicolumn{3}{c|}{3DFed paper}                                                                                                                                                                                                                                             \\ \cline{2-4} 
                                                                          & \multicolumn{1}{c|}{20}                                                                   & \multicolumn{1}{c|}{\begin{tabular}[c]{@{}c@{}}Not\\ reported*\end{tabular}}                      & 96                                                                           \\ \cline{2-4} 
                                                                          & \multicolumn{3}{c|}{Our replication study}                                                                                                                                                                                                                                   \\ \cline{2-4} 
                                                                          & \multicolumn{1}{c|}{20**}                                                                 & \multicolumn{1}{c|}{\textit{61.1}}                                                                & \textbf{0.56}                                                                \\
                                                                          & \multicolumn{1}{c|}{25}                                                                   & \multicolumn{1}{c|}{\textit{63.9}}                                                                & \textbf{0.55}                                                                \\
                                                                          & \multicolumn{1}{c|}{30}                                                                   & \multicolumn{1}{c|}{\textit{65.2}}                                                                & \textbf{0.55}                                                                \\
                                                                          & \multicolumn{1}{c|}{40}                                                                   & \multicolumn{1}{c|}{\textit{66.8}}                                                                & \textbf{0.55}                                                                \\ \cline{1-1} \hline
\end{tabular}
\end{table}

\vspace{-0.5em}
\section{Replication Study of 3DFed}
\label{sec:app-3dfed}
\verfiveChanged{To ensure consistency with \thrdfed work, we used the exact source code provided by \thrdfed authors \cite{3dfedCode} and thoroughly followed the experimental setup described in 3DFed paper \cite{li2023sp}. Every aspect, from hyper-parameter selection to training regime configuration, was carefully implemented. We repeated each experiment a minimum of 50 times to ensure robustness and reliability. To introduce variability, we employed various random seed values in each experiment. This rigorous process was applied to four distinct pre-trained models, each reaching different convergence levels based on the number of training epochs. Our experimental results, summarized in Tab. \ref{tab:rep-3dfed}, primarily focus on two key metrics: Backdoor Accuracy (BA) and Main-task Accuracy (MA). The findings undeniably establish the effectiveness of Flame in mitigating 3DFed attacks. As the global model approaches convergence, a significant reduction in BA is observed, indicating that the BA closely resemble those obtained through random guesses.} 

\FloatBarrier
\vfill
\ifminorrevision
\onecolumn
\input{11_minor_supp}
\else
\fi

\end{document}